\documentclass[%
 aip,
 amsmath,amssymb,
 reprint,
]{revtex4-1}

\usepackage{graphicx}
\usepackage{dcolumn}
\usepackage{bm}

\usepackage[utf8]{inputenc}
\usepackage[T1]{fontenc}
\usepackage{mathptmx}

\usepackage{xcolor}
\usepackage{dcolumn}

\newcommand{\kB}{k_\mathrm{B}}
\newcommand{\mbr}{\mathbf{r}}

\newcommand{\Zdu}{{Z_2^{:}}}  
\newcommand{\Zdd}{{Z_2^{|}}}  
\newcommand{\Ztu}{{Z_3^\therefore}} 
\newcommand{\Ztd}{{Z_3^{\cdot |}}}  
\newcommand{\Ztt}{{Z_3^{\triangle}}}  
\newcommand{\Zqd}{{Z_4^{:|}}} 
\newcommand{\Zqt}{{Z_4^{\cdot \triangle}}} 
\newcommand{\Zqq}{{Z_4^{||}}} 
\newcommand{\Zqc}{{Z_4^{\square}}} 

\DeclareUnicodeCharacter{2212}{-}

\begin{document}


\title[Fourth virial coefficient of helium isotopes]{Path-integral
  calculation of the fourth virial coefficient of 
  helium isotopes}

\author{Giovanni Garberoglio}
\email{garberoglio@ectstar.eu}
 \affiliation{
   European Centre for Theoretical Studies in Nuclear Physics and
        Related Areas (FBK-ECT*), Trento, I-38123, Italy}
\affiliation{Trento Institute for Fundamental Physics and Applications
  (TIFPA-INFN), Trento, I-38123, Italy}

\author{Allan H. Harvey}
\email{allan.harvey@nist.gov}

\affiliation{Applied Chemicals and Materials Division, National Institute of
  Standards and Technology, Boulder, CO 80305, USA}

\date{\today}

\begin{abstract}
We use the path-integral Monte Carlo (PIMC) method and state-of-the-art two-body and three-body potentials to calculate the fourth virial coefficients $D(T)$ of $^4$He and $^3$He as functions of temperature from 2.6~K to 2000~K.
We derive expressions for the contributions of exchange effects due to the bosonic or fermionic nature of the helium isotope; these effects have been omitted from previous calculations.
The exchange effects are relatively insignificant for $^4$He at the temperatures considered, but for $^3$He they are necessary for quantitative accuracy below about 4~K. 
Our results are consistent with previous theoretical work (and with some of the limited and scattered experimental data) for $^4$He; for $^3$He there are no experimental values and this work provides the first values of $D(T)$ calculated at this level.
The uncertainty of the results depends on the statistical uncertainty of the PIMC calculation, the estimated effect of omitting four-body and higher terms in the potential energy, and the uncertainty contribution propagated from the uncertainty of the potentials.
At low temperatures, the uncertainty is dominated by the statistical uncertainty of the PIMC calculations, while at high temperatures the uncertainties related to the three-body potential and to omitted higher-order contributions become dominant.
\end{abstract}

\maketitle

\section{Introduction}

State-of-the-art metrology for temperature and pressure increasingly relies on properties of helium calculated from first principles.
Helium is unique among noble gases in that its small number of electrons allows the pair interaction between atoms to be computed almost exactly; the nonadditive three-body interaction can also be calculated accurately.
Statistical mechanics (sometimes along with other atomic properties such as the polarizability that can also be accurately calculated) can then provide properties of helium gas more accurately than they can be measured.
Examples of this approach include modern acoustic gas thermometry,\cite{Moldover_2014} dielectric-constant gas thermometry,\cite{Gaiser_2015} refractive-index gas thermometry,\cite{Rourke_2019} and the recent development of a primary pressure standard based on dielectric measurements of helium.\cite{Gaiser20}

Helium also has the advantage of remaining in the gaseous state to low temperatures, making it the only feasible gas for metrology below about 25~K.
Most metrology uses natural helium which is predominantly the $^4$He isotope, but the rare isotope $^3$He may be used at very low temperatures due to its even lower liquefaction temperatures.

The most important quantities in these applications are the virial coefficients that define the equation of state of a gas of molar density $\rho$ at
temperature $T$ in an expansion around the ideal-gas (zero-density) limit,
\begin{equation}
  \frac{p}{\rho R T} = 1 + B(T) \rho + C(T) \rho^2 + D(T) \rho^3 + \cdots,
  \label{eq:virial}
\end{equation}
where $p$ is the pressure and $R$ is the molar gas constant.  
The second virial coefficient $B(T)$ depends on the interaction between two molecules, the third virial coefficient $C(T)$ depends on interactions among three molecules, and so on.
While for many substances the most accurate values of these coefficients are those obtained from careful analysis of density data, for small molecules first-principles calculations may be able to obtain smaller uncertainties than experiment.
Calculation of virial coefficients from intermolecular potentials will be described in the next section; here we note that, due to the low mass of helium, classical virial coefficient calculations are insufficient and quantum effects must be included (including exchange effects at very low temperatures).

For helium, in 2012 Cencek \textit{et al.}\cite{u2_2010} reported values of $B(T)$ of unprecedented accuracy calculated from a pair potential that incorporated higher-order effects (adiabatic correction to the Born-Oppenheimer approximation, relativistic effects, quantum electrodynamics).
The potential was further improved in 2017 by Przybytek \textit{et al.},\cite{Przybytek_2017} and still further in 2020 by Czachorowski \textit{et al.}\cite{u2_2020}
The 2020 work reports $B(T)$ for both $^4$He and $^3$He with uncertainties 5-10 times smaller than the uncertainties obtained by Cencek \textit{et al.}\cite{u2_2010}
This accuracy results both from the highly accurate pair potential and from the fact that an exact quantum calculation of $B(T)$ is possible via a phase-shift method.\cite{Hirschfelder:54}

For the third virial coefficient $C(T)$, no exact solution is known, but the fully quantum result can be approached numerically with the path-integral Monte Carlo (PIMC) method.
The most accurate first-principles values of $C(T)$ come from Garberoglio \textit{et al.},\cite{Garberoglio11,Garberoglio11err,Garberoglio2011,improvederr}
who used the same pair potential as Cencek \textit{et al.}\cite{u2_2010} and the three-body potential reported by Cencek \textit{et al.}\cite{FCI}  They found that it was necessary to account for non-Boltzmann statistics (exchange) below approximately 5~K for $^4$He and 6~K for $^3$He.\cite{Garberoglio11,Garberoglio11err}
They also estimated the uncertainty of $C(T)$, which primarily arose from the uncertainty of the three-body potential and (at low temperatures) from the statistical uncertainty of the PIMC calculations.
These results, at least in the Boltzmann case, were confirmed
independently by the group of Kofke.\cite{Shaul12,Kofke19}

For many purposes, knowledge of $B(T)$ and $C(T)$ is sufficient.
However, at higher pressures, terms containing $D(T)$ become significant.  Such terms contributed to the uncertainty budgets of a recent pressure standard\cite{Gaiser20} and recent single-pressure refractive-index gas thermometry measurements at cryogenic temperatures.\cite{Gao_2020}
Efforts to calculate $D(T)$ for helium that include quantum effects have been rare. 
Garberoglio performed approximate calculations above 100~K using a centroid-based
method.\cite{Garberoglio2012} 
The group of Kofke came the closest to a rigorous calculation,\cite{Shaul12,Kofke19} employing PIMC (considering only Boltzmann statistics) to calculate $D(T)$ for $^4$He from 2.6~K to 1000~K based on accurate pair\cite{u2_2010} and three-body\cite{FCI} potentials.  However, no uncertainties were reported apart from the statistical uncertainty of the PIMC calculations, and the results of Garberoglio and Harvey\cite{Garberoglio11,Garberoglio11err} for $C(T)$ suggest that the neglect of exchange effects will cause these Boltzmann results to be in error at the lowest temperatures.

In this work, we use state-of-the-art pair\cite{u2_2020} and three-body\cite{FCI} potentials to compute $D(T)$, fully incorporating exchange effects in our PIMC calculations.  We also provide the first rigorous calculations of $D(T)$ for $^3$He, and provide uncertainty estimates that include not only statistical uncertainty but also the contribution of uncertainties in the potentials.

\section{Virial coefficients and exchange effects}

In Eq.~(\ref{eq:virial}), the virial coefficients are known from statistical
mechanics~\cite{Hirschfelder:54, Hill-intro, Hill}
and depend on the $N$-body partition functions $Q_N(V,T)$ according to
\begin{eqnarray}
  \frac{B(T)}{N_\mathrm{A}} &=& -\frac{Z_2 - Z_1^2}{2V} \label{eq:B} \\
  \frac{C(T)}{{N_\mathrm{A}}^2} &=& \frac{(Z_2 - Z_1^2)^2}{V^2} - \frac{Z_3 -
    3Z_2Z_1 + 2  
    Z_1^3}{3V} \label{eq:C} \\
  \frac{D(T)}{{N_\mathrm{A}}^3} &=& -\frac{Z_4 - 4 Z_3 Z_1 - 3 Z_2^2 + 12 Z_2 Z_1^2 - 6 Z_1^4}{8V} +
  \nonumber \\
  & & \frac{3(Z_2-Z_1^2) (Z_3 - 3 Z_2 Z_1 + 2 Z_1^3)}{2 V^2} -
  \frac{5 (Z_2-Z_1^2)^3}{2 V^3},  \label{eq:D}
\end{eqnarray}
where $N_\mathrm{A}$ is the Avogadro constant and the auxiliary functions $Z_N$ are
defined as
\begin{equation}
  \frac{Z_N}{N!} = \frac{Q_N(V,T) ~ V^N}{Q_1(V,T)^N}.
\label{eq:ZN}  
\end{equation}

The appearance of powers of the Avogadro constant in the definition of
virial coefficients is due to the fact that Eq.~(\ref{eq:virial}) has been
written in terms of molar quantities, as the virial coefficients are
usually reported; for the sake of conciseness we will omit these factors in
subsequent formulae for $B$, $C$, and $D$ with the understanding that they must be applied as in Eqs. (\ref{eq:B})-(\ref{eq:D}) to produce the virial coefficients in molar units.

In general, the partition functions $Q_N(V,T)$ are given by
\begin{eqnarray}
    Q_N(V,T) &=& {\sum_i}' \langle i | \mathrm{e}^{-\beta H_N} | i \rangle
  \label{eq:QN_1} \\
  &=& \frac{1}{N!} \sum_{i,\sigma} \langle i | \mathrm{e}^{-\beta H_N} P_\sigma
  |i\rangle,
  \label{eq:QN}
\end{eqnarray}
where $H_N = K_N + V_N$ is the Hamiltonian of an $N$ particle system, where
$K_N$ is the total kinetic energy and $V_N$ the total potential energy of $N$
particles. The non-additive part of the $N$-body potential
will be denoted $u_N$ instead, so that $V_2 = u_2$, $V_3 = u_3 +
\sum_{i<j}^3 u_2(i,j)$, and $V_4 = u_4 + \sum_{i<j<k}^4 u_3(i,j,k) +
\sum_{i<j}^4 u_2(i,j)$.
The primed sum in Eq.~(\ref{eq:QN_1}) is over the many-body states
$|i\rangle$ with the correct symmetry upon particle exchange in the case of
bosonic or fermionic particles.
Equation~(\ref{eq:QN}) is an equivalent expression for the partition
function, where the sum is over a complete set of many-body states
irrespective of the symmetry upon exchange and $\sigma$ are the
permutations of $N$ objects. $P_\sigma$ is an operator performing the
permutation in the Hilbert space, which is multiplied by the sign of the
permutation in the case of fermions.

In general, not all the degrees of freedom of the state of the system that
we have denoted by $|i\rangle$ appear in the Hamiltonian $H_N$. We will
denote by $x$ the set of the degrees of freedom appearing in $H_N$ (atomic
coordinates, in our case) and with $s$ the other ones (nuclear spins), so
that we can write $|i\rangle = |s\rangle |x\rangle$.

In the following we will use Eqs.~(\ref{eq:B}--\ref{eq:QN}) to derive
explicit expressions for the quantum statistical contributions to virial
coefficients that apply at low temperatures. The analogous formula for the
second virial coefficient has been known for a long time,~\cite{Boyd69} but
its derivation will be useful to fix the notation used in the remainder of
the paper. In the case of the third virial coefficient, we will provide
derivation of the formulae reported in
Refs.~\onlinecite{Garberoglio11,Garberoglio11err}. The fourth virial
coefficient is newly developed in this work.

\subsection{The second virial coefficient}

Let us begin with the calculation of $Q_1$. There is only one (trivial)
permutation appearing in Eq.~(\ref{eq:QN}), and we will denote it by
$P^\cdot$. Performing the sum over the states we obtain~\cite{Garberoglio11}
\begin{equation}
  Q_1(V,T) = n \frac{V}{\Lambda^3},
\end{equation}
where $n$ is the number of internal states of the atoms we are considering
and $\Lambda = h / \sqrt{2 \pi m \kB T}$ is the de~Broglie thermal
wavelength of a particle of mass $m$, which here can be either the
mass of the ${}^4$He atom or the mass of the ${}^3$He atom. 
For helium isotopologues, $n = 2I+1$ where $I$ is the nuclear spin state;
$I=0$ for ${}^4$He (so $n = 1$) and $I=1/2$ for ${}^3$He ($n = 2$).  
Notice that $Z_1 = V$.

In the evaluation of $Q_2(V,T)$, we must consider two permutations. The
first one is the identity, which we will denote by $P^{:}$, whereas the
other exchanges the labels of the two particles and we will denote it by
$P^{|}$. The latter permutation is odd, and hence its contribution is
weighted with a $+$ sign in the case of bosons and a $-$ sign in the case
of fermions.
We have
\begin{widetext}
\begin{eqnarray}
  Q_2(V,T) &=& \frac{1}{2} \left[
    \left( \sum_{s_1,s_2} \langle s_1, s_2 | P^{:} | s_1, s_2 \rangle \right)
    \langle x_1, x_2 | \mathrm{e}^{-\beta H_2} P^{:} | x_1, x_2 \rangle \pm
    \left( \sum_{s_1,s_2} \langle s_1, s_2 | P^{|} | s_1, s_2 \rangle \right)
    \langle x_1, x_2 | \mathrm{e}^{-\beta H_2} P^{|} | x_1, x_2 \rangle
    \right] \nonumber \\
  &=& \frac{1}{2} \left[
    n^2
    \langle x_1, x_2 | \mathrm{e}^{-\beta H_2} P^{:} | x_1, x_2 \rangle \pm
    n
    \langle x_1, x_2 | \mathrm{e}^{-\beta H_2} P^{|} | x_1, x_2 \rangle
    \right], \\
  &\equiv& \frac{1}{2} \left[
    n^2 Q_2^{:}(V,T) \pm n Q_2^{|}(V,T) \label{eq:Q2}  \right],
  \end{eqnarray}
\end{widetext}
where the last equation defines the Boltzmann and exchanged partition
function, given by $Q_2^{:}$ and $Q_2^{|}$, respectively.  The factor $n^2$
in front of $Q_2^:$ comes from the fact that $P^{:}$ is the identity and
the weight $n$ in front of $Q_2^|$ from the fact that 
$$
\sum_{s_1,s_2} \langle s_1, s_2 | P^{|} | s_1, s_2 \rangle =
\sum_{s_1,s_2} \langle s_1, s_2 | s_2, s_1 \rangle =
\sum_{s_1,s_2} (\delta_{s_1, s_2})^2 = n.
$$

As detailed in our previous work,~\cite{Garberoglio2011,Garberoglio11} the
Boltzmann partition function  can be evaluated in the path-integral
framework and its final expression is equivalent to the partition function of a
system where each of the two quantum particles is replaced by a classical
ring polymer with $P$ monomers (the equivalence being exact in the
$P\to\infty$ limit).
The theory describes the form of the probability for a ring-polymer
configuration $F(\Delta\mbr_1, \ldots, \Delta\mbr_{P-1};
m,P,T)$,~\cite{Garberoglio11} where $\Delta \mbr_i = \mbr_{i+1}-\mbr_i$ is
the separation between the position of a bead and the subsequent one. Since
the distribution $F$ is translationally invariant, it is convenient to assume that $\mbr_1$ is the
null vector. Notice that the distance between the last bead and the first
one is given by the condition of having a closed polymer, hence $\mbr_P -
\mbr_1 = - \sum_{i=1}^{P-1} \Delta \mbr_i$.

Analogously, the exchanged partition function is equivalent to the
partition function of a {\em single} ring polymer of $2P$ monomers.
Finally, from Eqs.~(\ref{eq:Q2}), (\ref{eq:ZN}), and (\ref{eq:B}), one
obtains, for any value of the nuclear spin $I$,
\begin{eqnarray}
  B(T) &=& B_\mathrm{Boltz}(T) + \frac{(-1)^{2I}}{2I+1} B_\mathrm{xc}(T) \\
  B_\mathrm{Boltz}(T) &=& -\frac{1}{2V} 
  \left( Z_2^{:} - V^2 \right) \\
  B_\mathrm{xc}(T) &=& -\frac{Z_2^{|}}{2V}  
\end{eqnarray}
where, denoting by $\overline{V_2^{:}}$ and $\overline{V_2^{|}}$ the
potential energies of the equivalent classical systems for the identity and
the swap permutations, one has
\begin{eqnarray}
  \frac{Z_2^{:}}{V} &=& \int \mathrm{d}\mbr \left\langle \mathrm{e}^{-\beta \overline{V_2^{:}}}
  \right\rangle \label{eq:Z2B} \\
  \frac{Z_2^{|}}{V} &=& \frac{\Lambda^3}{2^{3/2}} \left\langle
  \mathrm{e}^{-\beta \overline{V_2^|}} \right\rangle. \label{eq:Z2xc}
\end{eqnarray}
The average in Eq.~(\ref{eq:Z2B}) is on the configurations of two ring polymers of $P$
beads, whereas the average in Eq.~(\ref{eq:Z2xc}) is on just one ring
polymer of $2P$ beads (and mass $m/2$). In any case, the configurations 
are sampled according to the distribution $F$ described above. The average
potentials are defined as
\begin{eqnarray}
  \overline{V_2^{:}}(\mbr) &=&
  \frac{1}{P} \sum_{i=1}^P u_2\left(\left|\mbr + \mbr_i^{(1)} -
  \mbr_i^{(2)} \right| \right)  \label{eq:V2B} \\
  \overline{V_2^|} &=&
  \frac{1}{P} \sum_{i=1}^{P}  u_2\left(\left|\mbr_i^{(3)} - \mbr_{i+P}^{(3)} \right| \right),
  \label{eq:V2xc}
\end{eqnarray}
where $\mbr_i^{(1)}$ and $\mbr_i^{(2)}$ in Eq.~(\ref{eq:V2B}) denote the
coordinates of two independent ring polymers of $P$ beads each, and
$\mbr_i^{(3)}$  in Eq.~(\ref{eq:V2xc}) denotes the coordinates of a ring polymer
of $2P$ beads.
The factor $\Lambda^3 / 2^{3/2}$ in Eq.~(\ref{eq:Z2xc}) comes from how the action of
the permutation operator $P^{|}$ on the product of the probability
distributions $F$ of two $P$-monomer polymers produces the probability
distribution of a $2P$-monomer coalesced polymer.~\cite{Garberoglio11}

\subsection{The third virial coefficient}

The same considerations leading to Eq.~(\ref{eq:Q2}) can be applied to the
three-particle partition function appearing in the expression of $C(T)$.
In this case, the six permutations of three particles can be conveniently
divided into three subsets. The first set includes only the identity permutation,
whose representation we will denote as $P^\therefore$. The second set includes
the three permutations that swap two particles (which are odd in
character), whose
representation will be denoted by $P^{\cdot |}$, whereas the third set
includes the two remaining cyclic permutations, which are even, and will be
denoted by $P^\triangle$. The analogous expression to Eq.~(\ref{eq:Q2}) is then
\begin{equation}
  Q_3(V,T) = \frac{1}{3!} \left(
  n^3 Q_3^{\therefore} \pm 3 n^2 Q_3^{\cdot |} + 2 n Q_3^\triangle
  \right).
  \label{eq:Q3}
\end{equation}

Using Eqs.~(\ref{eq:Q3}) and (\ref{eq:Q2}) together with (\ref{eq:ZN}), one can
write Eq.~(\ref{eq:C}) as
\begin{eqnarray}
  C(T) &=& C_\mathrm{Boltz} + \frac{(-1)^{2I}}{2I+1} C_\mathrm{odd}(T) +
  \frac{C_\mathrm{even}}{(2I+1)^2} \\
  C_\mathrm{Boltz} &=& \frac{(Z_2^{:} - V^2)^2}{V^2} -
  \frac{Z_3^{\therefore} - 3V Z_2^{:} + 2
    V^3}{3V}  \\
  C_\mathrm{odd}(T) &=& -\frac{1}{V} \left( \Ztd - V \Zdd \right) + \frac{2}{V^2}
  \left(\Zdu - V^2 \right) \Zdd \\
  C_\mathrm{even}(T) &=& \frac{1}{V^2} \left( \Zdd^2 - \frac{2}{3}V \Ztt \right),
\end{eqnarray}
where, denoting as $\overline{V_3^{\cdot |}}$ the total three-body energy
when two particles are coalesced into a single ring polymer
and  by $\overline{V_3^{\triangle}}$ the total three-body  energy when
three particles are coalesced in a single ring polymer,~\cite{Garberoglio11} we have
\begin{eqnarray}
  \frac{\Ztu}{V} &=& 
  \int \mathrm{d}\mbr_1 \mathrm{d}\mbr_2
  \left\langle
  \exp\left(-\beta \overline{V_3^\therefore}(\mbr_1, \mbr_2)\right)
  \right\rangle \label{eq:Ztu} \\
  \frac{Z_3^{\cdot |}}{V} &=& \frac{\Lambda^3}{2^{3/2}}
  \int \mathrm{d}\mbr
  \left\langle
  \exp\left(-\beta \overline{V_3^{\cdot |}}(\mbr)\right) \right\rangle \\
    \frac{Z_3^{\triangle}}{V} &=&
  \frac{\Lambda^6}{3^{3/2}}
  \left\langle
  \exp\left(-\beta \overline{V_3^{\triangle}}\right)
  \right\rangle,
\end{eqnarray}
where, analogously to Eqs.~(\ref{eq:V2B}) and (\ref{eq:V2xc}), we have
defined
\begin{eqnarray}
  \overline{V_3^\therefore}(\mbr_1, \mbr_2) &=&
    \frac{1}{P} \sum_{i=1}^P V_3(\mbr_1 + \mbr_i^{(1)}, \mbr_2 + \mbr_i^{(2)}, \mbr_{i+P}^{(3)}) \label{eq:V33} \\
  \overline{V_3^{\cdot |}}(\mbr) &=&
    \frac{1}{P} \sum_{i=1}^P V_3(\mbr_i^{(1)}, \mbr + \mbr_i^{(4)}, \mbr +
    \mbr_{i+P}^{(4)}) \label{eq:V34} \\
    \overline{V_3^{\triangle}} &=&
    \frac{1}{P} \sum_{i=1}^P
    V_3(\mbr_i^{(5)}, \mbr_{i+P}^{(5)}, \mbr_{i+2P}^{(5)}). \label{eq:V35}
\end{eqnarray}
In Eqs. (\ref{eq:V33})--(\ref{eq:V35}), $\mbr_i^{(k)}$ for $k=1,2,3$ are the coordinates of independent $P$-bead
ring polymers, $\mbr_i^{(4)}$ are the coordinates of a ring polymer with
$2P$ beads and mass $m/2$, and $\mbr_i^{(5)}$ denote the coordinates of a
ring polymer with $3P$ beads and mass $m/3$.
Note that we have slightly changed the notation from
Refs.~\onlinecite{Garberoglio11,Garberoglio11err}, by collecting together all the
terms with an odd or even character upon particle exchange.

\subsection{The fourth virial coefficient}

The permutation group of 4 particles has an even richer structure. For our
purposes, it is sufficient to recall the presence of the following subsets:
\begin{itemize}
\item The identity element, whose representation we will denote as
  $P^{::}$. In this case the sum over the internal states gives a factor
  $n^4$.
\item The swapping of a single pair. This subset has odd parity, and
  includes 6 elements. Its representation will be denoted by $P^{:|}$. The
  sum over the internal states results in a factor $n^3$.
\item The cyclic permutation on subsets of 3 particles. This subset has an
  even parity and includes 8 elements. Its representation will be denoted
  by $P^{\cdot \triangle}$. The sum on the internal states produces a
  factor of $n^2$.
\item The swapping of two distinct pairs. This subset has even parity and
  includes 3 elements. Its representation will be denoted
  by $P^{||}$. Also in this case the sum over the internal states produces
  a factor $n^2$.
\item The 6 remaining permutations all produce a single ring polymer. This subset has odd parity
  and includes, for example, the cyclic permutations. Its representation
  will be denoted by  $P^\square$, and the sum over the internal states
  produces a factor of $n$.
\end{itemize}

We can then write the 4-particle partition function as
\begin{equation}
  Q_4(V,T) = \frac{1}{4!} \left(
  n^4 Q_4^{::} \pm 6 n^3 Q_4^{:|} + 8 n^2 Q_4^{\cdot \triangle} + 3 n^2
  Q_4^{||}
  \pm 6 n Q_4^\square
  \right),
\end{equation}
and the expression for  $D(T)$ turns out to be, after lengthy but
straightforward calculations,

\begin{widetext}
\begin{eqnarray}
  D(T) &=& D_\mathrm{Boltz}(T) +  D_\mathrm{xc}(T) \label{eq:Ddef} \\
  D_\mathrm{Boltz}(T) &=& -\frac{Z_4^{::} - 4 V Z_3^{\therefore} - 3
    (Z_2^{:})^2 + 12 V^2 Z_2^{:} - 6 V^4}{8V} +
  \frac{3(Z_2^{:}-V^2) (Z_3^{\therefore} - 3 V Z_2^{:} + 2 V^3)}{2 V^2} -
  \frac{5 (Z_2^{:}-V^2)^3}{2 V^3} \label{eq:DBoltz}\\
  D_\mathrm{xc}(T) &=& \frac{(-1)^{2I}}{2I+1} D_\mathrm{o1}(T) +
  \frac{1}{(2I+1)^2} D_\mathrm{e1}(T) + \frac{(-1)^{2I}}{(2I+1)^3} D_\mathrm{o2}(T) \label{eq:Dxc} \\
     &=& D_\mathrm{o1}(T) + D_\mathrm{e1}(T) + D_\mathrm{o2}(T)~~[\mathrm{for~^4He}] \label{eq:Dxc-4} \\
     &=& \frac{-1}{2} D_\mathrm{o1}(T) + \frac{1}{4} D_\mathrm{e1}(T) + \frac{-1}{8}D_\mathrm{o2}(T)~~[\mathrm{for~^3He}] \label{eq:Dxc-3} \\
  D_\mathrm{o1}(T) &=& -\frac{3}{4V} \left(\Zqd - 2 V \Ztd - \Zdu \Zdd + 2 V^2 \Zdd
  \right) + \frac{9}{2V^2} (\Zdu -V^2) (\Ztd - V \Zdd) + \nonumber \\
  &  & \frac{3}{2V^2}
  \left( \Ztu - 3 V \Zdu + 2 V^3 \right) \Zdd -\frac{15}{2V^3} \left(
  \Zdu - V^2 \right)^2  \Zdd \label{eq:Do1} \\
  D_\mathrm{e1}(T) &=& - \frac{(\Zqt - V \Ztt)}{V} -
  \frac{3 (\Zqq - \Zdd^2)}{8V} +
  \frac{3 (\Zdu - V^2) \Ztt}{V^2} +
  \frac{9 \Zdd (\Ztd - V \Zdd)}{2V^2} - 
  \frac{15 \left( \Zdu - V^2 \right) \Zdd^2}{2V^3} \label{eq:De1} \\
  D_\mathrm{o2}(T) &=& -\frac{3}{4V} \Zqc + \frac{3}{V^2} \Zdd \Ztt -
  \frac{5}{2V^3} \Zdd^3,
  \label{eq:Do2}
\end{eqnarray}
\end{widetext}
where, denoting again with $\overline{V_4^\sigma}$ the total four-body
potential energy for the equivalent classical system obtained by applying
the permutation $\sigma$, we have defined
\begin{eqnarray}
  \frac{Z_4^{::}}{V} &=&
  \int \mathrm{d}\mbr_1 \mathrm{d}\mbr_2 \mathrm{d}\mbr_3
  \left\langle
  \exp\left(-\beta \overline{V_4^{::}}\right)
  \right\rangle \label{eq:Z4B} \\
  \frac{Z_4^{:|}}{V} &=&
  \frac{\Lambda^3}{2^{3/2}}
  \int \mathrm{d}\mbr_1 \mathrm{d}\mbr_2
  \left\langle
  \exp\left(-\beta \overline{V_4^{:|}}\right)
  \right\rangle \\
  \frac{Z_4^{||}}{V} &=&
  \frac{\Lambda^6}{8}
  \int \mathrm{d}\mbr
  \left\langle
  \exp\left(-\beta \overline{V_4^{||}}\right)
  \right\rangle \\
  \frac{Z_4^{\cdot \triangle}}{V} &=&
  \frac{\Lambda^6}{3^{3/2}}
  \int \mathrm{d}\mbr
  \left\langle
  \exp\left(-\beta \overline{V_4^{\cdot\triangle}}\right)
  \right\rangle \\
  \frac{Z_4^{\square}}{V} &=&
  \frac{\Lambda^9}{8}
  \left\langle
  \exp\left(-\beta \overline{V_4^{\square}}\right)
  \right\rangle, \label{eq:Zsquare}
\end{eqnarray}
with
\begin{eqnarray}
  \overline{V_4^{::}}(\mbr_1, \mbr_2, \mbr_3) &=& \frac{1}{P} \sum_{i=1}^P
  V_4\left(\mbr_1 + \mbr_i^{(1)}, \mbr_2 + \mbr_i^{(2)},  \mbr_3 + \mbr_i^{(3)},
  \mbr_i^{(4)}\right) \label{eq:V44} \\
  \overline{V_4^{:|}}(\mbr_1, \mbr_2) &=& \frac{1}{P} \sum_{i=1}^P
  V_4\left(\mbr_1 + \mbr_i^{(1)}, \mbr_2 + \mbr_i^{(2)},
  \mbr_i^{(5)}, \mbr_{i+P}^{(5)}\right) \\
  \overline{V_4^{||}}(\mbr) &=& \frac{1}{P} \sum_{i=1}^P
  V_4\left(\mbr_i^{(5)}, \mbr_{i+P}^{(5)}, \mbr + \mbr_i^{(6)},  \mbr +
  \mbr_{i+P}^{(6)}\right) \\  
  \overline{V_4^{\cdot \triangle}}(\mbr) &=& \frac{1}{P} \sum_{i=1}^P
  V_4\left(\mbr + \mbr_i^{(1)}, \mbr_i^{(7)},  \mbr_{i+P}^{(7)},
  \mbr_{i+2P}^{(7)}\right) \\
  \overline{V_4^\square} &=& \frac{1}{P} \sum_{i=1}^P
  V_4\left(\mbr_i^{(8)}, \mbr_{i+P}^{(8)}, \mbr_{i+2P}^{(8)}, \mbr_{i+3P}^{(8)}\right). \label{eq:V48}
\end{eqnarray}
In Eqs.~(\ref{eq:V44})--(\ref{eq:V48}), $\mbr_i^{(k)}$ with $k=1,2,3,4$ are the
coordinates of independent $P$-bead
ring polymers, $\mbr_i^{(k)}$ with $k=5,6$ are the coordinates of
independent $2P$-bead
ring polymers and mass $m/2$, $\mbr_i^{(7)}$ are the coordinates of a $3P$-bead
ring polymer and mass $m/3$, and $\mbr_i^{(8)}$ are the coordinates of a $4P$-bead
ring polymer and mass $m/4$.

\section{Numerical calculations}

In order to find the optimal number of monomers $P$ as a function of temperature in the ring-polymer representation of the quantum problem, we
reanalyzed highly accurate calculations of $B(T)$ to find the values $P(T)$ above which the calculated $B$ did not change within about one part in $10^4$. 
We found that
convergence in the investigated range of temperature was assured by
choosing $P$ equal to $\mathrm{nint}(4 + 620 / (T/1~\mathrm{K})^{0.7})$ in
the case of ${}^4$He and $\mathrm{nint}(4 + 770 /
(T/1~\mathrm{K})^{0.7})$ in the case of ${}^3$He, where $\mathrm{nint}(x)$
denotes the nearest integer to $x$. With this choice, the values
of $P$ are close to what we have used in our previous work for $T > 10$~K,
but optimize the utilization of numerical resources in the low-temperature
regime where degeneracy is important.
We validated this choice of $P$ by repeating the calculations with the number
of beads doubled for $10$~K, $120$~K, and $1000$~K and observing that the
results agreed within their statistical uncertainties.

In our calculations of $D(T)$, we performed the multidimensional
integrations in Eqs.~(\ref{eq:Ddef})--(\ref{eq:Zsquare}) using the
parallel implementation of the VEGAS algorithm.\cite{vegas2,Kreckel97}
The expression for $D_\mathrm{Boltz}$ in Eq.~(\ref{eq:DBoltz}) can be
written as an integration over the three positions of the first bead of the
ring polymers corresponding to three atoms (the fourth has its first bead
fixed at the origin of the coordinate system due to translational
invariance). This nine-dimensional integration can further be reduced to
six dimensions due to of rotational invariance. Notice that all the terms
in Eq.~(\ref{eq:DBoltz}) can be written as multidimensional integrals on
three vector positions; for example the term $V Z_3^\therefore$ is of this
form since two vector positions come from the definition of
$Z_3^\therefore$ in Eq.~(\ref{eq:Ztu}) and the third comes from writing $V
= \int d^3\mbr$. Analogous considerations are valid for the exchange
contributions to $D(T)$. In this case, ring polymers
corresponding to distinguishable particles coalesce in larger polymers, and
the number of coordinates used to identify their configuration is
correspondingly smaller. For example, in all the terms contributing to
$D_\mathrm{o1}(T)$ we have to deal with three ring polymers instead of four and
we just need two vectors to identify them by the position of the first
bead; rotational invariance further reduces the number of independent
coordinates to three.

The six-dimensional integrations leading to $D_\mathrm{Boltz}$ were
performed using $4 \times 10^6$ Monte Carlo calls, the three-dimensional
integrations leading to $D_\mathrm{o1}$ using $5 \times 10^5$ calls, and the
one-dimensional integrations leading to $D_\mathrm{e1}$ using $5000$ calls.
The averages appearing in Eqs.~(\ref{eq:Z4B})--(\ref{eq:Zsquare}) were
evaluated using independent ring polymers generated anew using the
prescription of Levy~\cite{Levy54,Jordan-Fosdick68} for each of the
coordinates where the integrand is required by the integration procedure.
The number of ring polymers used depends on the specific contribution to
$D(T)$: we used 16 in the case of $D_\mathrm{Boltz}$, 128 for
$D_\mathrm{o1}$ and $D_\mathrm{e1}$, and $3\times 10^6$ for the evaluation
of $D_\mathrm{o2}$.
Following Kofke and coworkers,~\cite{Shaul12,Kofke19} we found it
convenient to evaluate separately the two-body contribution to the various
components of $D(T)$ and the difference leading to the full calculation
involving the three-body potential.

We used the very recent two-body potential by Czachorowski {\em et
  al.}~\cite{u2_2020} and the three-body potential by Cencek {\em  et
  al.}~\cite{FCI} To the best of our knowledge, no four-body potential is
available for helium, so in this work we set it to zero.
The additional uncertainty due to this assumption is discussed in Sec.~\ref{sec:unc-4}.

The evaluation of $D(T)$ using the procedure outlined above is quite CPU-intensive. 
At the lowest temperature investigated here, where the
calculations are more demanding, we needed roughly 20~000 CPU-hours on a
modern 2.5~GHz processor, half of which are dedicated to the evaluation of
the exchange contributions. The requirements are roughly inversely
proportional to the temperature, hence we needed 4000 CPU-hours at
$T$~=~10~K, 650 CPU-hours at 120~K, and so on.

\section{Estimation of uncertainties}\label{sec:unc}

\subsection{PIMC statistical uncertainty}

The statistical uncertainty of the PIMC calculations was evaluated
as the standard error of the mean of a set of independent calculations. The
number of independent calculations that we employed varied according to the
component of $D(T)$ that was calculated. We used 
8 for the Boltzmann component at high temperature and up to 40 at the
lowest temperatures, 24 for $D_\mathrm{o1}$, 64 for
$D_\mathrm{e1}$ and $D_\mathrm{o2}$.
The VEGAS algorithm produces its own estimation of standard uncertainty,
which we verified was in very good agreement with the estimate based on
independent calculations.

\subsection{Uncertainty propagated from potentials}
The usual way to evaluate the contribution to the overall uncertainty of
virial coefficients due to the potentials is to perform the calculation
with perturbed two- and three-body potentials (where the size of the perturbation is the estimated uncertainty of the potential) and take the difference.~\cite{Garberoglio2009b,Garberoglio2011,Cencek2013}
Given the considerable computing requirements for the evaluation of $D(T)$,
we developed a more efficient and more accurate way to propagate the uncertainty from the
potentials to the fourth virial coefficient, starting from the functional
derivative of Eq.~(\ref{eq:D}) with respect to the $k$-body irreducible
potentials $u_k(\mbr_1,\ldots,\mbr_k)$ that $D(T)$ depends on.
We begin by noticing that in the classical limit, the functions $Z_N$ become
\begin{equation}
  Z_N = \int \mathrm{e}^{-\beta V_N(\mbr_1,\ldots,\mbr_N)} \prod_{i=1}^N \mathrm{d}^3\mbr_i, 
\label{eq:ZN_class}
\end{equation}
and that the variation of the $n$-th virial coefficient, $B_n$, can be
written as
\begin{equation}
  \delta B_n^{(k)} = \int \frac{\delta B_n}{\delta u_k(\mbr_1, \ldots,
    \mbr_k)} \delta u_k(\mbr_1,\ldots,\mbr_k)~ 
  \prod_{i=1}^k \mathrm{d}^3\mbr_i,  
\end{equation}
where $\delta B_n / \delta u_k$ is the functional derivative of the virial
coefficient with respect to the $k$-body irreducible potential.
We will use this equation as a starting point to estimate how the
uncertainty of the potentials is propagated to the uncertainty in the
virial coefficients, assuming that $\delta u_k$ is the expanded uncertainty
estimated during the {ab initio} derivation of $u_k$. Since we know of no other
constraint on the possible variations of $u_k$ (such as, for example,
estimates of the uncertainties on the forces, $\delta u'_k$, or
higher-order derivatives of $u_k$)
we will neglect this aspect in the following. In order to
compensate for this approximation, we will further adopt some
conservative choices, as detailed in the discussion below. From
Eq.~(\ref{eq:ZN_class}) one has
\begin{equation}
  \frac{\delta{Z_2}}{\delta u_2(\mbr_1, \mbr_2)} = -\beta
  \exp\left[-\beta u_2(\mbr_1, \mbr_2)\right]
  \label{eq:dZ2du2}
\end{equation}
\begin{equation}
  \frac{\delta{Z_3}}{\delta u_2(\mbr_1, \mbr_2)} = - 3 \beta
  \int \exp\left[-\beta V_3(\mbr_1, \mbr_2, \mbr_3)\right] ~ \mathrm{d}\mbr_3
  \end{equation}
\begin{equation}
  \frac{\delta{Z_3}}{\delta u_3(\mbr_1, \mbr_2, \mbr_3)} = - \beta
  \exp\left[-\beta V_3(\mbr_1, \mbr_2, \mbr_3)\right]
  \end{equation}
\begin{equation}
  \frac{\delta{Z_4}}{\delta u_2(\mbr_1, \mbr_2)} = -6 \beta
  \int 
  \exp\left[-\beta V_4(\mbr_1, \ldots , \mbr_4)\right] ~ \mathrm{d}\mbr_3 \mathrm{d}\mbr_4
  \end{equation}
\begin{equation}
  \frac{\delta{Z_4}}{\delta u_3(\mbr_1, \mbr_2, \mbr_3)} = -4 \beta
  \int
  \exp\left[-\beta V_4(\mbr_1, \ldots, \mbr_4)\right] ~ \mathrm{d}\mbr_4
  \end{equation}
\begin{equation}
  \frac{\delta{Z_4}}{\delta u_4(\mbr_1, \mbr_2, \mbr_3, \mbr_4)} = -\beta 
  \exp\left[-\beta V_4(\mbr_1, \ldots, \mbr_4)\right],
  \label{eq:dZ4du4}
\end{equation}
enabling evaluation of $\delta D$ by functional differentiation of
Eq.~(\ref{eq:D}). Actually, one can evaluate the variation of $D(T)$ with
respect to the two- and three-body potentials separately; using as an example
the variation due to a perturbation of the pair potential, 
an expression is obtained of the form
\begin{equation}
  \delta D^{(2)}(T) = \int \delta u_2(\mbr_1, \mbr_2) \frac{\delta D}{\delta
    u_2(\mbr_1, \mbr_2)} ~ \mathrm{d}\mbr_1 \mathrm{d}\mbr_2.
\label{eq:dD2_naive}
\end{equation}

A naive evaluation of the uncertainty using (the absolute value of)
Eq.~(\ref{eq:dD2_naive}) when only $\delta u_2$ is assumed to be at any point
the absolute value of the estimated uncertainty due to the pair potential
is at best a lower bound on the actual uncertainty on $D(T)$. Actually, the
other factor in the integrand, that is the functional
derivative $\delta D/ \delta u_2$, is a function that has negative and
positive regions whose relative importance changes with temperature. Our
calculations show that the estimated uncertainty using
Eq.~(\ref{eq:dD2_naive}) would cross zero somewhere around $T=30$~K.
Conversely, a more conservative estimate (possibly resulting in an upper bound) to the propagated
uncertainty from the potential to the fourth virial coefficient can be
obtained by using the absolute value of the integrand in
Eq.~(\ref{eq:dD2_naive}).
Since we are considering potentials with uncertainties, the actual value of
$\delta D/\delta u_2$ is bounded by those calculated with the most positive
($u_2^+ = u_2 + \delta u_2$) or the most negative ($u_2^- = u_2 - \delta
u_2$) pair potential. We decided to adopt the conservative choice of using,
for each configuration in the integrand, the largest value of $|\delta D/\delta
u_2|$ between those calculated with the modified potentials $u_2^+$ and
$u_2^-$. With this provision, the final expression of the propagated
uncertainty due to the $k$-body potential is
\begin{equation}
  \delta D^{(k)}(T) = \int \delta u_k
  \left| \frac{\delta D}{\delta u_k}
  \right|_\mathrm{max} ~ \prod_{i=1}^4 \mathrm{d}^3\mbr_i.
  \label{eq:dDk}
\end{equation}
We considered the uncertainties associated to the potentials as expanded
uncertainties with coverage factor $k=2$, so we obtained the standard
uncertainties by dividing by $2$ the values obtained using
Eq.~(\ref{eq:dDk}). The standard uncertainties from the pair and three-body
potentials were then added in quadrature to obtain the standard
uncertainty due to the imperfect knowledge of the potential-energy surfaces.

The final formulae that we obtain are:
\begin{widetext}
  \begin{eqnarray}
    \delta D^{(2)}(T) &=& \frac{\beta}{8 V} \int
    \delta u_2(1,2) \left|
    6 \mathrm{e}^{-\beta V_4(1,2,3,4)} - 12 \mathrm{e}^{-\beta V_3(1,2,3)} - 6 \mathrm{e}^{-\beta
      (V_2(1,2) + V_2(3,4))} + 12 \mathrm{e}^{-\beta V_2(1,2)} \right. \nonumber \\
    & & -12 \mathrm{e}^{-\beta V_2(1,2)} \left(
    \mathrm{e}^{-\beta V_3(2,3,4)} - 3 \mathrm{e}^{-\beta V_2(3,4)} + 2 \right)
    -36
    \left(
    \mathrm{e}^{-\beta V_3(1,2,3)} - \mathrm{e}^{-\beta V_2(1,2)}
    \right)\left[ \mathrm{e}^{-\beta V_2(3,4)} -1 \right]
    \nonumber \\
        & & \left.
        + 60 \mathrm{e}^{-\beta V_2(1,2)} \left( \mathrm{e}^{-\beta V_2(3,4)}-1\right)^2
      \right|
    \prod_{i=1}^4 \mathrm{d}^3\mbr_i, \label{eq:dDu2}
  \end{eqnarray}
  \begin{equation}
    \delta D^{(3)}(T) = \frac{\beta}{8 V} \int
    \delta u_3(1,2,3) \left| 4 \mathrm{e}^{-\beta V_4(1,2,3,4)} - 4 \mathrm{e}^{-\beta
      V_3(1,2,3)} + 12 \mathrm{e}^{-\beta V_3(1,2,3)} \left( \mathrm{e}^{-\beta
      V_2(3,4)}-1\right)
    \right|     \prod_{i=1}^4 \mathrm{d}^3\mbr_i, \label{eq:dDu3}
  \end{equation}
\end{widetext}
where in actual numerical integration it can be useful to consider that
the final result must be invariant under any permutation of the four particles'
coordinates so that, for example, one can write $4 \mathrm{e}^{-\beta V_3(1,2,3)} =
\mathrm{e}^{-\beta V_3(1,2,3)} + \mathrm{e}^{-\beta V_3(2,3,4)} +
\mathrm{e}^{-\beta V_3(3,4,1)} + \mathrm{e}^{-\beta V_3(4,1,2)}$. An
analogous expression has been used to evaluate the terms involving the pair potential.
Our final estimate for the propagated standard uncertainty due to the potential is
then
\begin{equation}
  u_V(D) = \frac{1}{2} \sqrt{\left(\delta D^{(2)}(T)\right)^2 +
    \left(\delta D^{(3)}(T)\right)^2}.
  \label{eq:uV}
\end{equation}

Given the quantum nature of helium, we decided to use in the calculation of
the uncertainty (\ref{eq:uV}) the Feynman--Hibbs (FH) approach. This method
takes into account quantum effects by using modified potentials for
calculating the partition functions, and hence the $Z_i$. In this
framework, one can think of the potentials appearing in
Eqs.~(\ref{eq:dZ2du2})--(\ref{eq:dZ4du4}) as being already given in the FH
form.  Since the largest contribution to $D(T)$ comes from the pair
potential, we used the FH approach only for this case, keeping the
classical form of the three-body potential when evaluating
Eq.~(\ref{eq:dDk}). Additionally, we found it convenient to use the
fourth-order FH semiclassical pair potential, given
by~\cite{FH4}
\begin{eqnarray}
u_\mathrm{FH4}(r) &=& u_2(r) +
\frac{\beta \hbar^2}{12 m} \left( u_2''(r)  +\frac{2 u_2'(r)}{r}\right) +
\nonumber \\
& & 
\frac{1}{2}  \left( \frac{\beta \hbar^2}{12 m} \right)^2
\left( u_2''''(r) + \frac{4 u_2'''(r)}{r} \right).
\label{eq:FH4}
\end{eqnarray}
In Eq.~(\ref{eq:FH4}), the number of primes indicates the order of the
derivative which, in our calculations, have been evaluated numerically
using the central-difference formulae with a grid spacing $\delta r = 10^{-3}$~\AA.
We found that this approach provided very good estimates (that is, within
10\%) for the uncertainty of
$B(T)$ and $C(T)$, when
compared with the traditional way of calculating this
contribution,~\cite{u2_2010,Garberoglio2009b,Garberoglio2011,Cencek2013} down to
roughly $T=4$~K, corresponding to the onset of quantum exchange effects,
which are not taken into account in this procedure. Below this threshold,
the semiclassical estimation is likely to produce an upper bound on the actual
uncertainty, since using the quadratic FH potential in
Eqs.~(\ref{eq:dDu2}) and (\ref{eq:dDu3}) produces a
larger estimate of the uncertainty than the quartic used here.

At the lowest temperatures, we checked whether the approximations used to
evaluate $u_V(D)$ were still good by running calculations with perturbed
potentials and taking one fourth of the difference of the values of $D(T)$
obtained in this way, as in our previous work.~\cite{Garberoglio2009b,Garberoglio2011,Cencek2013}
Since the statistical uncertainty is large in this regime, we took the
conservative approach of adding one fourth of the combined statistical uncertainties of
$D(T)$ with the perturbed potentials to generate an estimate of $u_V(D)$.
We found this estimate to be compatible with the one obtained
using the procedure described in this section in the case of ${}^4$He: we
obtained an estimate of $u_V(D, T=2.6~\mathrm{K}) =
14500$~cm${}^9~$mol${}^{-3}$ from PIMC calculations, to be compared with the
value $15000$~cm${}^9$/mol${}^3$ obtained from the novel approach, and
reported in Table~\ref{tab:DHe4}.

In the case of ${}^3$He, the semiclassical estimation of $u_V(D)$ produces a
value of $2 \times 10^5$ cm${}^9$~mol${}^{-3}$ at $T=2.6$~K which is four times as large as the
value of $D(T)$, hence the approach described in this section is definitely
questionable at the lowest temperatures. Analogously to the case of
${}^4$He, we performed calculations with the perturbed potentials, finding
a good agreement above $T=5$~K. For lower temperatures, we report as
$u_V(D)$ the values obtained using the difference of the PIMC simulation
with perturbed potentials combined with the statistical uncertainty.

The way of estimating the uncertainty described here is much faster than
taking the difference between the values of $D(T)$ calculated with
perturbed potentials. In that case, one has to converge the calculation so
that the statistical uncertainty is smaller than the potential uncertainty;
the use of Eq.~(\ref{eq:uV}) provides directly the uncertainty and hence
can be evaluated with less computational effort. As an example, converging
$D(T)$ calculated using perturbed potentials to a statistical uncertainty
smaller than the potential uncertainty requires 1.5 hours of
CPU time (with a semiclassical approach), whereas the evaluation of
Eq.~(\ref{eq:uV}) takes only ten seconds on the same hardware.

\subsection{Uncertainty due to four-body potential}\label{sec:unc-4}

We performed these calculations with high-accuracy pair\cite{u2_2020} and
three-body\cite{FCI} potentials.  While this truncation of the many-body
expansion is rigorous for the third virial coefficient, the fourth virial
coefficient is affected by any four-body nonadditivity that might exist.
Unfortunately, there have been few attempts to calculate four-body
interactions for helium, and to our knowledge no four-body potential has
been developed.  Our analysis of this uncertainty contribution must
therefore involve some guesswork and approximations.

Because helium is not very polarizable, the multibody forces are weak and
the multibody expansion should converge quickly.  This is reflected in the
behavior of $C(T)$, where almost all of the quantity is given by the pair
potential and the contribution of the three-body potential is on the order
of 1\% to 2\%.\cite{Garberoglio2009b} We find similar behavior for $D(T)$,
where the three-body effect is much smaller than the two-body contribution at most temperatures.
At high temperatures, however, the relative
contribution of three-body effects becomes large (on the order of 30\% at
2000~K); this is probably an artifact of the absolute value of $D$ becoming
small.
Because of the apparent rapid convergence of the multibody
expansion, a rough estimate for the effect of four-body
forces might be 10\% of the size of the three-body effect.


In order to be more quantitative, we considered the four-body
potential developed by Bade in the framework of the Drude model of
dispersion.~\cite{Bade1,Bade3} This model takes into account only
long-range dispersion, so that it results in a pair
potential with a $r^{-6}$ dependence and in the case of three-body
forces it reproduces the Axilrod--Teller form. It has been used to estimate
the contribution of four-body forces to $D(T)$ in the cases of neon and
argon.~\cite{WiebkeNe,WiebkeAr}
Bade's potential depends on
the single-atom polarizability $\alpha_1$, which is known both experimentally
and theoretically with high precision for He,~\cite{Gaiser18,Puchalski20} and an unknown
parameter -- named $\hbar \omega_0$ or $V$ in the original papers -- that we fixed from
the knowledge of the coefficient of $r^{-6}$ term in the most recent pair
potential for He.~\cite{u2_2020}
With this choice, Bade's model produces a value for the Axilrod--Teller
parameter for He that is within 4\% of the actual value.\cite{Thakkar_1981}
We then proceeded to evaluate the contribution to $D(T)$ from this
four-body potential, using a semiclassical approach with the 4th-order
FH expression for the pair potential.
As might be expected based on the convergence of the many-body expansion, this contribution is on the order of 10\% of the three-body contribution, except near 5~K where the three-body contribution crosses zero.

However, Bade's model is purely for dispersion interactions; it does not take into account the repulsive part of the
four-body potential.
For $C(T)$, it is known that the Axilrod-Teller three-body dispersion effect is only accurate at low temperatures;
the repulsive nonadditivity has the opposite sign and becomes a larger contribution to $C(T)$ above approximately 170~K.\cite{Garberoglio2009b}
The magnitude of this repulsive contribution is similar to the size of the Axilrod-Teller contribution,
so similar behavior might be expected for the contribution of repulsive nonadditivity to $D(T)$.
We therefore estimated
an uncertainty contribution due to the missing four-body potential as the maximum of
15\% of the absolute value of the three-body contribution (interpolated at those temperatures
where we did not calculate it) and the absolute value of the dispersion contribution estimated from
Bade's potential. 
We consider this an expanded uncertainty with coverage factor $k = 2$.
The corresponding standard uncertainty is reported in
Table~\ref{tab:DHe4} for ${}^4$He and in Table~\ref{tab:DHe3} for ${}^3$He
in the column labeled $u_4(D)$.

\section{Results and Discussion}

\subsection{Calculated $D(T)$}

We report the results of our calculations for ${}^4$He in
Tables~\ref{tab:DHe4} and \ref{tab:DxcHe4}, where we also compare our
results with the values reported by the group of Kofke.~\cite{Shaul12,Kofke19}
In general, we obtain very good agreement for the Boltzmann
contribution to $D(T)$, although the statistical uncertainties of our
calculations are generally larger. 
In order to produce Boltzmann values of lower uncertainty based on all data, we combined their results
with ours at the temperatures where both studies reported results, with statistical weighting according to the uncertainty from each source:
\begin{eqnarray}
  D &=& \frac{\frac{D_\mathrm{TW}}{u(D_\mathrm{TW})^2} +
      \frac{D_\mathrm{KG}}{u(D_\mathrm{KG})^2}}
  {u(D_\mathrm{TW})^{-2} + u(D_\mathrm{KG})^{-2}} \\
    u(D) &=& \sqrt{\frac{1}{u(D_\mathrm{TW})^{-2} + u(D_\mathrm{KG})^{-2}}},
\end{eqnarray}
where the subscript TW indicates the values calculated in this work, and
the subscript KG indicates the value from Kofke's group.
While the work from the Kofke group used a different pair potential, the effect
of this difference on $D(T)$ is negligible compared to the statistical
uncertainty of the calculation, making it legitimate to combine the results
in this way.

At temperatures above about 8~K, the uncertainty
budget is dominated by the uncertainty due to the imperfect knowledge of
the potentials.
At the lower end of this temperature range, inspection of the contributions to $u_V(D)$ shows that this
is mainly due to the uncertainty of the three-body potential. 
Above about 20~K, the uncertainty is dominated by $u_4(D)$, the uncertainty due to possible four-body interactions.
At the lowest temperatures, the statistical uncertainty from the PIMC calculations (primarily the calculation of $D_\mathrm{Boltz}$)
becomes a sizable contribution to the overall uncertainty budget.

The effect of the exchange terms on $D(T)$ is minimal, at least in the case
of ${}^4$He. Inspection of Table~\ref{tab:DHe4} shows that the values of
$D_\mathrm{xc}(T)$ have an uncertainty comparable to their
absolute value. The breakdown of all the contributions, reported in
Table~\ref{tab:DxcHe4}, shows that $D_\mathrm{xc}(T)$ is obtained as the sum
of terms with opposite signs and comparable magnitude, resulting in
consistent cancellations.

The situation for ${}^3$He, reported in Tables~\ref{tab:DHe3} and
\ref{tab:DxcHe3}, is qualitatively similar to that of the heavier isotope,
with two notable differences. First, the fourth virial coefficient
increases with decreasing temperature in the whole range considered,
different from the case of ${}^4$He where a maximum is observed around $T = 4.5$~K. 
Second, the fermionic nature of ${}^3$He is such that the
various contributions to $D_\mathrm{xc}(T)$ in Eq.~(\ref{eq:Dxc-3}) are all positive, resulting
in a net positive value of the exchange terms in the whole range
considered.
Because the group of Kofke did not report calculations for ${}^3$He,\cite{Shaul12,Kofke19} 
the values of $D_\mathrm{Boltz}(T)$ in Table~\ref{tab:DHe3} are only those resulting from our calculations.

\begin{table*}
    \caption{The fourth virial coefficient of ${}^4$He and its
      contributions: $D_\mathrm{Boltz,TW}(T)$, Boltzmann contribution
      calculated in this work;
      $D_\mathrm{Boltz,KG}(T)$, Boltzmann contribution from
      Ref.~\onlinecite{Shaul12}; $D_\mathrm{Boltz}(T)$, combined
      Boltzmann values from this work and Ref.~\onlinecite{Shaul12};
      $D_\mathrm{xc}(T)$ exchange contribution; $u_V(D)$,
      propagated uncertainty from the pair and three-body potential;
      $u_4(D)$, estimated uncertainty due to the neglect of the
      four-body potential;
      $D(T)$, final values.
      The numbers in parentheses report standard
      uncertainty on the last digits. In the last column, the combined expanded
      uncertainty of $D(T)$ is reported at $k=2$ coverage.}
\label{tab:DHe4}   
  \begin{ruledtabular}
  \begin{tabular}{d|c|c|c|c|c|c|c|c}
    \multicolumn{1}{c|}{Temperature} &
    \multicolumn{1}{c|}{$D_\mathrm{Boltz,TW}(T)$} &
    \multicolumn{1}{c|}{$D_\mathrm{Boltz,KG}(T)$} &
    \multicolumn{1}{c|}{$D_\mathrm{Boltz}(T)$} &        
    \multicolumn{1}{c|}{$D_\mathrm{xc}(T)$} &
    \multicolumn{1}{c|}{$u_V(D)$} &
    \multicolumn{1}{c|}{$u_4(D)$} &
    \multicolumn{1}{c|}{$D(T)$} &
    \multicolumn{1}{c}{$U(D)$} \\
    \multicolumn{1}{c|}{(K)} &
    \multicolumn{1}{c|}{(cm${}^9$~mol${}^{-3}$)} &
    \multicolumn{1}{c|}{(cm${}^9$~mol${}^{-3}$)} &
    \multicolumn{1}{c|}{(cm${}^9$~mol${}^{-3}$)} &
    \multicolumn{1}{c|}{(cm${}^9$~mol${}^{-3}$)} &
    \multicolumn{1}{c|}{(cm${}^9$~mol${}^{-3}$)} &    
    \multicolumn{1}{c|}{(cm${}^9$~mol${}^{-3}$)} &
    \multicolumn{1}{c|}{(cm${}^9$~mol${}^{-3}$)} &
    \multicolumn{1}{c}{(cm${}^9$~mol${}^{-3}$)} \\
    \hline
2.6	&	-242000(12000)	&	-238000(8000)	&	-239000(7000)	&	-900(3000)	&	15000	&	1400	&	-240000	&	30000	\\
2.8	&			&	-120000(6000)	&	-120000(6000)	&	-1100(2300)	&	8000	&	900	&	-120000	&	20000	\\
3	&	-53000(6000)	&	-61000(4000)	&	-58000(3000)	&	700(1000)	&	4000	&	600	&	-60000	&	11000	\\
3.25	&			&	-17000(3000)	&	-17000(3000)	&	1100(600)	&	3000	&	300	&	-16000	&	8000	\\
3.5	&			&	6000(2000)	&	6000(2000)	&	-200(400)	&	1700	&	300	&	6000	&	5000	\\
3.75	&			&	12400(1500)	&	12000(1500)	&	0(300)	&	1100	&	200	&	12000	&	4000	\\
4	&	19000(2000)	&	19500(1100)	&	19400(1000)	&	160(80)	&	800	&	180	&	20000	&	3000	\\
4.25	&			&	20800(900)	&	20800(900)	&	50(120)	&	600	&	160	&	21000	&	2000	\\
4.5	&			&	22500(700)	&	22500(700)	&	54(80)	&	500	&	140	&	23000	&	1700	\\
4.75	&			&	21000(500)	&	21000(500)	&	-49(40)	&	400	&	120	&	21000	&	1300	\\
5	&	21100(900)	&	19200(400)	&	19500(400)	&	-17(20)	&	300	&	110	&	19500	&	1000	\\
5.25	&			&	17300(400)	&	17300(400)	&	-1(20)	&	300	&	100	&	17300	&	1000	\\
5.5	&			&	16100(300)	&	16100(300)	&	-22(12)	&	200	&	100	&	16100	&	800	\\
6	&	13800(500)	&	13200(200)	&	13300(190)	&	-4(4)	&	170	&	90	&	13300	&	500	\\
7	&	8800(300)	&	9100(130)	&	9000(120)	&	-1.3(13)	&	110	&	70	&	9000	&	300	\\
8	&	6200(200)	&			&	6200(200)	&			&	70	&	60	&	6200	&	500	\\
8.5	&			&	5730(60)	&	5730(60)	&			&	60	&	50	&	5700	&	200	\\
10	&	4100(100)	&	4100(30)	&	4100(30)	&			&	40	&	40	&	4100	&	140	\\
15	&	2880(40)	&			&	2880(40)	&			&	17	&	30	&	2900	&	110	\\
20	&	2768(16)	&	2782(3)	&	2781(3)	&			&	9	&	20	&	2780	&	50	\\
24.5561	&			&	2740.1(14)	&	2740.1(14)	&			&	6	&	18	&	2740	&	40	\\
30	&	2665(6)	&	2659.0(8)	&	2659.1(8)	&			&	4	&	15	&	2660	&	30	\\
40	&			&	2462.5(3)	&	2462.5(3)	&			&	3	&	11	&	2460	&	20	\\
50	&	2256.7(18)	&	2259.90(19)	&	2259.90(19)	&			&	2	&	9	&	2260	&	19	\\
63.15	&			&	2025.61(12)	&	2025.61(12)	&			&	1.4	&	8	&	2030	&	15	\\
80	&	1781.3(7)	&			&	1781.3(7)	&			&	1.1	&	6	&	1781	&	13	\\
83.15	&			&	1742.22(7)	&	1742.22(7)	&			&	1.1	&	6	&	1742	&	12	\\
103.15	&			&	1526.67(4)	&	1526.67(4)	&			&	0.9	&	5	&	1527	&	10	\\
120	&	1381.7(5)	&			&	1381.7(5)	&			&	0.8	&	4	&	1382	&	9	\\
123.15	&			&	1358.59(4)	&	1358.59(4)	&			&	0.8	&	4	&	1359	&	9	\\
143.15	&			&	1224.22(3)	&	1224.22(3)	&			&	0.7	&	4	&	1224	&	8	\\
173.15	&			&	1066.096(16)	&	1066.096(16)	&			&	0.6	&	3	&	1066	&	6	\\
200	&	955.4(4)	&			&	955.4(4)	&			&	0.6	&	3	&	955	&	6	\\
223.15	&			&	876.991(16)	&	876.991(16)	&			&	0.6	&	3	&	877	&	5	\\
250	&	800.1(3)	&			&	800.1(3)	&			&	0.5	&	2	&	800	&	5	\\
273.15	&			&	744.051(12)	&	744.051(12)	&			&	0.5	&	2	&	744	&	5	\\
273.16	&	743.5(3)	&			&	743.5(3)	&			&	0.5	&	2	&	744	&	5	\\
300	&	687.0(3)	&			&	687.0(3)	&			&	0.5	&	2	&	687	&	4	\\
323.15	&			&	645.170(10)	&	645.170(10)	&			&	0.5	&	2	&	645	&	4	\\
400	&	534.0(2)	&	534.115(7)	&	534.115(7)	&			&	0.5	&	2	&	534	&	4	\\
500	&	433.9(2)	&	434.029(5)	&	434.029(5)	&			&	0.4	&	2	&	434	&	4	\\
600	&			&	363.438(5)	&	363.438(5)	&			&	0.4	&	2	&	363	&	4	\\
700	&	310.80(18)	&	310.920(4)	&	310.920(4)	&			&	0.4	&	2	&	311	&	4	\\
800	&			&	270.314(4)	&	270.314(4)	&			&	0.4	&	1.9	&	270	&	4	\\
900	&			&	238.021(4)	&	238.021(4)	&			&	0.4	&	1.9	&	238	&	4	\\
1000	&	211.50(14)	&	211.702(3)	&	211.702(3)	&			&	0.4	&	1.9	&	212	&	4	\\
1500	&	130.43(10)	&			&	130.43(10)	&			&	0.3	&	1.7	&	130	&	4	\\
2000	&	89.14(8)	&			&	89.14(8)	&			&	0.3	&	1.6	&	89	&	3	
  \end{tabular}    
  \end{ruledtabular}
\end{table*}      

\begin{table*}
  \caption{The exchange contributions to $D(T)$ for ${}^4$He, see
      Eqs.~(\ref{eq:Ddef})--(\ref{eq:Do2}). Uncertainties here are $k=1$.}
  \label{tab:DxcHe4}       
  \begin{ruledtabular}
    \begin{tabular}{d|c|c|c|c}
     \multicolumn{1}{c|}{Temperature} &
     \multicolumn{1}{c|}{$D_\mathrm{o1}(T)$} &
     \multicolumn{1}{c|}{$D_\mathrm{e1}(T)$} &
     \multicolumn{1}{c|}{$D_\mathrm{o2}(T)$} &
     \multicolumn{1}{c}{$D_\mathrm{xc}(T)$} \\
     \multicolumn{1}{c|}{(K)} &
     (cm${}^9 $~mol${}^{-3}$) &
     (cm${}^9 $~mol${}^{-3}$) &
     (cm${}^9 $~mol${}^{-3}$) &    
     (cm${}^9 $~mol${}^{-3}$) \\
     \hline
2.6 &   -5716(2663)  &   11324(1297)  &   -6475(73)    &   -867(2963)  \\
2.8 &   -3163(2236)  &   5072(544)   &   -2989(23)    &   -1081(2301)  \\
3   &   -2960(887)   &   5118(449)   &   -1464(17)    &   695(995)   \\
3.25    &   -286(585)   &   1945(149)   &   -596(5) &   1062(603)   \\
3.5 &   -839(438)   &   856(84)    &   -258(3) &   -241(446)   \\
3.75    &   -491(298)   &   580(54)    &   -113.9(13)    &   -25(303)   \\
4   &   -198(74)    &   407(33)    &   -51.9(9) &   158(81)    \\
4.25    &   -77(116)   &   151(14)    &   -24.6(4) &   50(117)   \\
4.5 &   -46(75)    &   111(10)    &   -11.6(2) &   54(75)    \\
4.75    &   -91(40)    &   47(6) &   -5.64(11)    &   -49(40)    \\
5   &   -47(21)    &   32(3) &   -1.39(4) &   -17(21)    \\
5.25    &   -21(22)    &   21(2) &   -0.73(2) &   -1(22)    \\
5.5 &   -27(12)    &   7.9(15)    &   -2.85(7) &   -22(12)    \\
6   &   -5(4) &   1.3(3) &   -0.2(1) &   -4(4) \\
7   &   -1.4(13)    &   0.20(6) &   -0.016(1) &   -1.3(13)    
    \end{tabular}    
  \end{ruledtabular}
\end{table*}

\begin{table*}
    \caption{The fourth virial coefficient of ${}^3$He and its
      contributions. $D_\mathrm{xc}$ is defined in Eq.~(\ref{eq:Dxc}) and
      the other quantities are as in Table~\ref{tab:DHe4}. The
      expanded uncertainty in the last column is at $k=2$ coverage.}     
  \label{tab:DHe3}   
  \begin{ruledtabular}
  \begin{tabular}{d|c|c|c|c|c|c}
    \multicolumn{1}{c|}{Temperature} &
    \multicolumn{1}{c|}{$D_\mathrm{Boltz}(T)$} &
    \multicolumn{1}{c|}{$D_\mathrm{xc}(T)$} &
    \multicolumn{1}{c|}{$u_V(D)$} &
    \multicolumn{1}{c|}{$u_4(D)$} &    
    \multicolumn{1}{c|}{$D(T)$} &
    \multicolumn{1}{c}{$U(D)$} \\    
    \multicolumn{1}{c|}{(K)} &
    \multicolumn{1}{c|}{(cm${}^9 $~mol${}^{-3}$)} &
    \multicolumn{1}{c|}{(cm${}^9 $~mol${}^{-3}$)} &
    \multicolumn{1}{c|}{(cm${}^9 $~mol${}^{-3}$)} &    
    \multicolumn{1}{c|}{(cm${}^9 $~mol${}^{-3}$)} &
    \multicolumn{1}{c|}{(cm${}^9 $~mol${}^{-3}$)} &    
    \multicolumn{1}{c}{(cm${}^9  $~mol${}^{-3}$)}\\
    \hline
2.6	&	55000(15000)	&	20600(2200)	&	9000	&	1300	&	75000	&	30000	\\
3	&	35000(9000)	&	5100(1300)	&	4000	&	500	&	40000	&	19000	\\
4	&	22600(3000)	&	900(500)	&	1300	&	160	&	23500	&	7000	\\
5	&	14100(1500)	&	170(50)	&	1400	&	90	&	14300	&	4000	\\
6	&	8900(900)	&	60(30)	&	600	&	70	&	8900	&	2000	\\
7	&	7000(500)	&	10(8)	&	300	&	60	&	7000	&	1300	\\
8	&	5900(400)	&	1(2)	&	200	&	50	&	5900	&	900	\\
10	&	5020(180)	&	0.0(3)	&	100	&	40	&	5020	&	400	\\
15	&	3790(90)	&			&	30	&	30	&	3790	&	190	\\
20	&	3540(30)	&			&	16	&	20	&	3535	&	80	\\
30	&	3130(12)	&			&	7	&	14	&	3128	&	40	\\
50	&	2483(4)	&			&	2	&	9	&	2483	&	20	\\
80	&	1890.6(14)	&			&	1.2	&	6	&	1891	&	13	\\
120	&	1437.6(7)	&			&	0.8	&	4	&	1438	&	9	\\
200	&	979.2(3)	&			&	0.6	&	3	&	979	&	6	\\
250	&	816.5(3)	&			&	0.5	&	2	&	817	&	5	\\
273.16	&	757.6(3)	&			&	0.5	&	2	&	758	&	5	\\
300	&	699.3(3)	&			&	0.5	&	2	&	699	&	4	\\
400	&	541.1(2)	&			&	0.5	&	2	&	541	&	4	\\
500	&	438.61(17)	&			&	0.4	&	2	&	439	&	4	\\
700	&	313.22(16)	&			&	0.4	&	1.9	&	313	&	4	\\
1000	&	212.84(15)	&			&	0.4	&	1.9	&	213	&	4	\\
1500	&	130.88(5)	&			&	0.3	&	1.8	&	131	&	4	\\
2000	&	89.41(4)	&			&	0.3	&	1.6	&	89	&	3	
  \end{tabular}
  \end{ruledtabular}
\end{table*}    
  
\begin{table*}
    \caption{The exchange contributions to $D(T)$ for ${}^3$He, see
      Eqs.~(\ref{eq:Ddef})--(\ref{eq:Do2}). Uncertainties here are $k=1$.}
  \label{tab:DxcHe3}         
  \begin{ruledtabular}
    \begin{tabular}{d|c|c|c|c}
    \multicolumn{1}{c|}{Temperature} &
    \multicolumn{1}{c|}{$D_\mathrm{o1}(T)$} &
    \multicolumn{1}{c|}{$D_\mathrm{e1}(T)$} &
    \multicolumn{1}{c|}{$D_\mathrm{o2}(T)$} &
    \multicolumn{1}{c}{$D_\mathrm{xc}(T)$}  \\  
    \multicolumn{1}{c|}{(K)} &
    \multicolumn{1}{c|}{(cm${}^9 $~mol${}^{-3}$)} &
    \multicolumn{1}{c|}{(cm${}^9 $~mol${}^{-3}$)} &
    \multicolumn{1}{c|}{(cm${}^9 $~mol${}^{-3}$)} &    
    \multicolumn{1}{c}{(cm${}^9  $~mol${}^{-3}$)} \\
    \hline
2.6 &   -22257(2985)  &   27603(896)   &   -20809(51)    &   20631(2158)  \\
3   &   -3185(1759)  &   10980(328)   &   -6263(15)    &   5120(1255)  \\
4   &   -1044(692)   &   1281(42)    &   -414.4(13)    &   894(490)   \\
5   &   -230(75)    &   197(8) &   -38(2) &   169(53)    \\
6   &   -93(40)    &   39(2) &   -4.35(3) &   57(28)    \\
7   &   -17(12)    &   8(5) &   -0.602(7) &   10(8) \\
8   &   -0.3(34)    &   1.6(1) &   -0.091(2) &   0.6(24)    \\
10  &   0.0(4) &   0.02(1) &   -0.003(3) &   0.02(27)    
    \end{tabular}    
  \end{ruledtabular}
  \end{table*}

\subsection{Comparison with experimental data}
While the virial coefficients in Eq.~(\ref{eq:virial}) can be extracted from experimental measurements,
in practice this becomes quite difficult for the higher coefficients because they require extracting higher-order effects
(a third derivative in the case of $D$).  
Relatively high pressures are also necessary to reach densities where $D(T)$ is significant.

The Burnett method of systematic expansions between two vessels of similar size can reduce the uncertainties in measuring virial coefficients.
This method was used by Canfield and coworkers in a series of papers on helium and its mixtures with other gases, several of which reported virial coefficients up to the fourth for helium at various temperatures.\cite{Blancett_1970,Hall_1970a,Hall_1970b,Provine_1971}
None of these papers reported a full uncertainty analysis; when plotting these points we use the parameter uncertainties reported in the papers, which appear to be only the statistical uncertainties from their fits to their data.

Highly accurate measurements on helium were performed by McLinden and L\"{o}sch-Will\cite{McLinden_2007} with a two-sinker magnetic-suspension densimeter.
The same method was used by Moldover and McLinden,\cite{Moldover_2010} who reanalyzed both their data and the earlier measurements.\cite{McLinden_2007}
This analysis was constrained with \textit{ab initio} values of $B(T)$\cite{u2_2010} and $C(T)$,\cite{Garberoglio2009b} allowing them to reduce the uncertainty of $D$ on each isotherm by a factor of 6 compared to unconstrained analysis.

A single value of $D$ at 273.16~K can be extracted from the dielectric-constant gas thermometry experiments of Gaiser and Fellmuth.\cite{Gaiser_2019}
The result is 723(79)~$\mathrm{cm}^9$~$\mathrm{mol}^{-3}$.\cite{Gaiser_2021}

\begin{figure}
\includegraphics[width=0.95\linewidth]{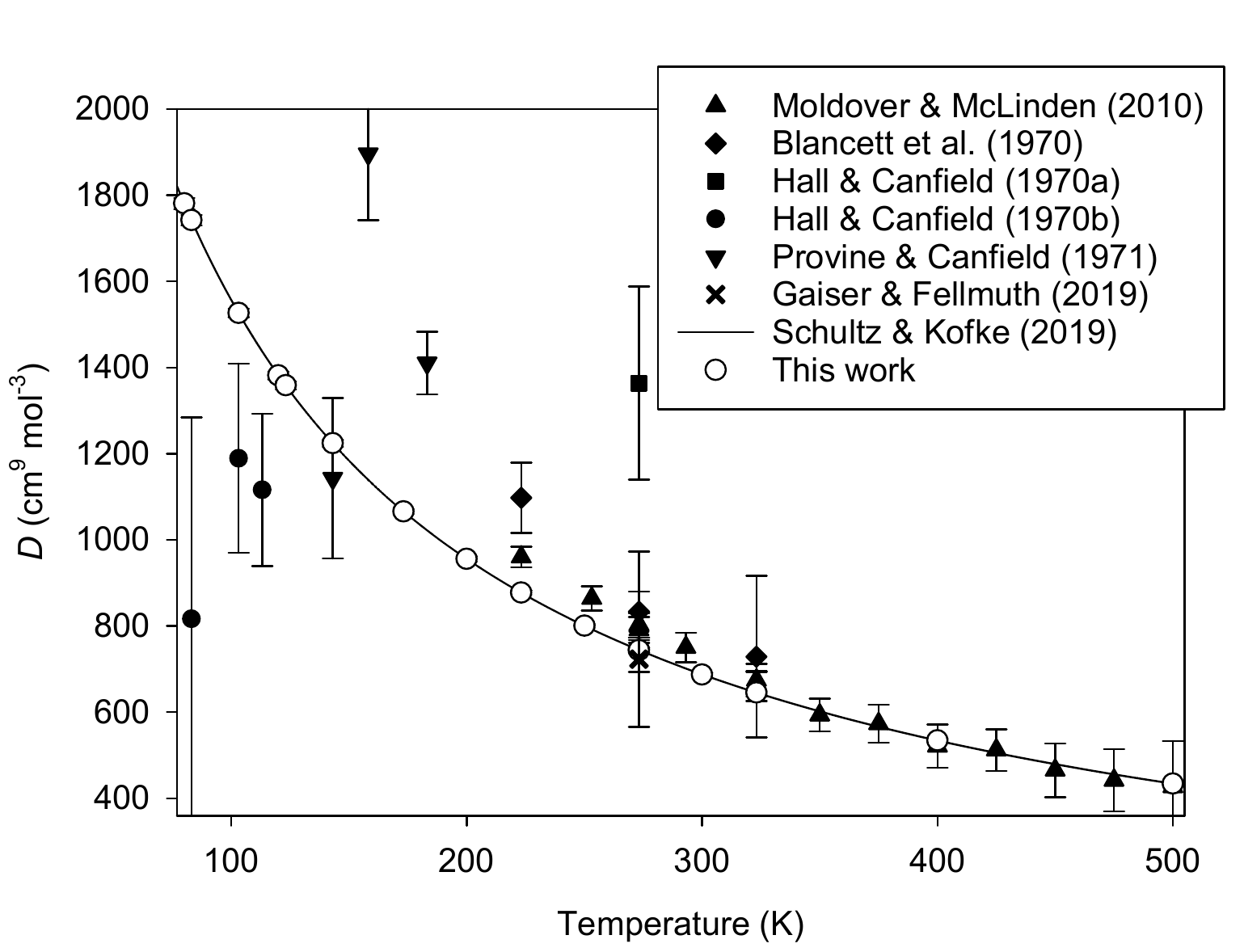}  
  \caption{Comparison of calculated values of $D(T)$ for ${}^4$He
    with experimental results\cite{Blancett_1970,Hall_1970a,Hall_1970b,Provine_1971,Moldover_2010,Gaiser_2021} and with correlation fitted to first-principles calculations by Schultz and Kofke.\cite{Kofke19}
    Error bars represent expanded uncertainties with coverage factor $k = 2$. Expanded uncertainties for this work are smaller than the size of the symbols; see Table~\ref{tab:DHe4}.  }
\label{fig:He4-data}  
\end{figure}

In Fig.~\ref{fig:He4-data}, our results from Table~\ref{tab:DHe4} are shown along with the experimental data for ${}^4$He.
The data from the Burnett experiments of 50 years ago\cite{Blancett_1970,Hall_1970a,Hall_1970b,Provine_1971} are in qualitative agreement with our results, but are too scattered for meaningful quantitative comparison.  The single dielectric-constant gas thermometry point of Gaiser and Fellmuth\cite{Gaiser_2019,Gaiser_2021} has a relatively large uncertainty, but agrees well with our results.
The results of Moldover and McLinden\cite{Moldover_2010} are in excellent agreement with our calculations above about 300~K. 
At lower temperatures, there is a disagreement which is small but outside the mutual uncertainties.
We note that the points of Moldover and McLinden that agree well are all from their ``2007 isotherms,'' while their points from the reanalyzed ``2005 isotherms'' measured by McLinden and L\"{o}sch-Will\cite{McLinden_2007} show a systematic offset (a differing trend between the two sets was also noted by Shaul \textit{et al.}\cite{Shaul_2012}). 
This may suggest an unrecognized problem with the 2005 experiments.
The 2007 isotherms were measured with new sinkers whose volumes were newly calibrated; it is plausible that there may have been a small error in calibration for the sinkers used in 2005.\cite{McLinden_2020}
Figure~\ref{fig:He4-data} also shows the equation that Schultz and Kofke\cite{Kofke19} fit to their results at 20~K and above, which were calculated using Boltzmann statistics only.  It is not surprising that this equation is in excellent agreement with our results, since exchange effects are negligible at these temperatures.

\subsection{Low-temperature behavior}
\begin{figure}
\includegraphics[width=0.95\linewidth]{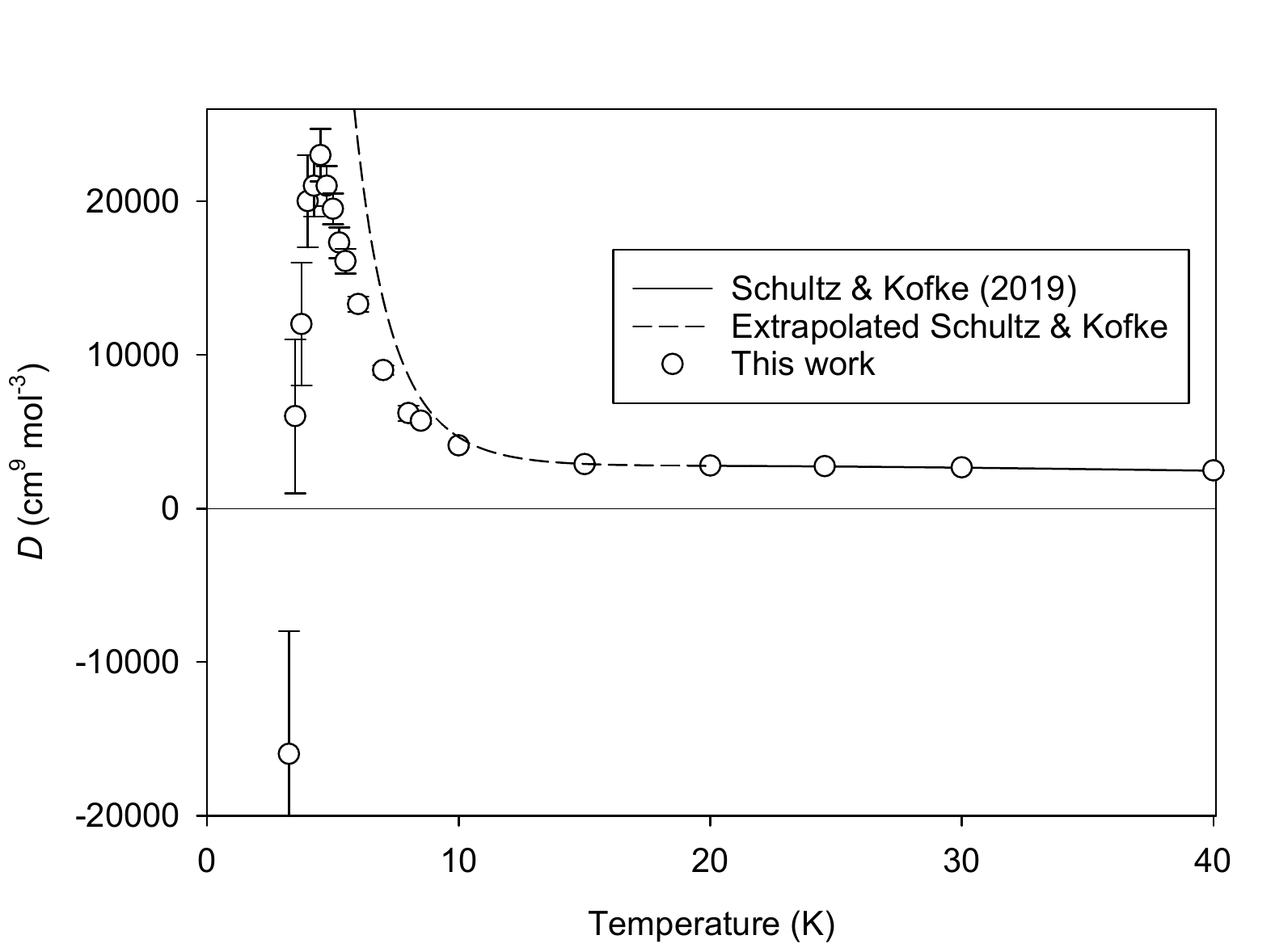}  
  \caption{Calculated values of $D(T)$ for ${}^4$He at low temperatures.
    The correlation fitted to first-principles calculations by Schultz and Kofke\cite{Kofke19} is extrapolated below 20~K.
    Error bars represent expanded uncertainties with coverage factor $k = 2$, and are smaller than the symbol size at higher temperatures.  }
\label{fig:He4-lowT}  
\end{figure}

\begin{figure}
\includegraphics[width=0.95\linewidth]{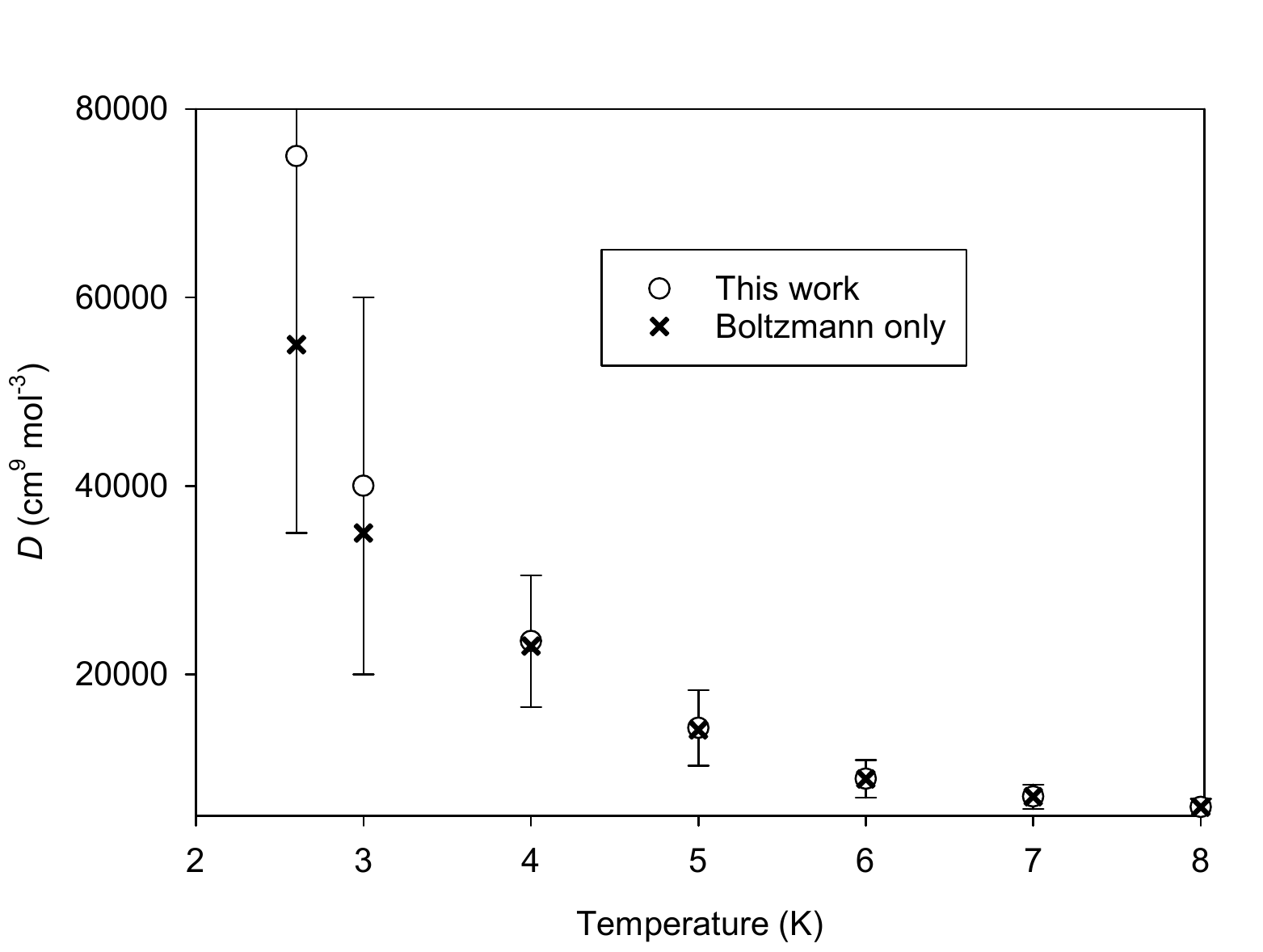}  
  \caption{Calculated values of $D(T)$ for ${}^3$He at low temperatures, also showing results from Boltzmann statistics alone.
    Error bars represent expanded uncertainties with coverage factor $k = 2$.  }
\label{fig:He3-lowT}  
\end{figure}

Since one of the novel features of this work is the incorporation of exchange effects that become important at low temperature, in Figs.~\ref{fig:He4-lowT} and \ref{fig:He3-lowT} we show low-temperature results for ${}^4$He and ${}^3$He, respectively.
No low-temperature experimental data are available for comparison in either case.

In Fig.~\ref{fig:He4-lowT}, we see that $D(T)$ for ${}^4$He goes through a maximum near 4.5~K, turning sharply negative at the lowest temperatures (the three lowest temperatures in Table~\ref{tab:DHe4} are not plotted; they would be far below the bottom of the graph).
We do not show the effect of exchange, since as discussed above the exchange contributions are relatively small compared to their uncertainty due to terms of opposite sign.
The correlation that Schultz and Kofke\cite{Kofke19} fitted to their Boltzmann results is in excellent agreement with the results obtained here down to its lower temperature limit of 20~K.  It extrapolates well down to 15~K, but becomes increasingly inaccurate when extrapolated further.

Figure~\ref{fig:He3-lowT} shows the low-temperature results for ${}^3$He.
No maximum is evident in the temperature range investigated.
We also show the values of $D(T)$ obtained when only Boltzmann statistics are considered.  Incorporation of exchange effects is necessary for quantitative accuracy below roughly 4~K.
The significant exchange effects on $D(T)$ for ${}^3$He, compared to a smaller effect for ${}^4$He due to competing terms, is similar to the behavior found in our previous work for $C(T)$.\cite{Garberoglio11,Garberoglio11err}

\section{Conclusions}

The values of $D(T)$ presented in Table~\ref{tab:DHe4} (for $^4$He) and Table~\ref{tab:DHe3} (for $^3$He) represent the first such values calculated from the current state-of-the-art pair and three-body potentials.  They are also the first to include exchange effects, and the first to include complete uncertainty estimates.  For ${}^3$He, Table~\ref{tab:DHe3} presents the first fully quantum calculations of $D(T)$.

Our ${}^4$He calculations at the level of Boltzmann statistics confirm previous calculations (with a slightly different two-body potential) from the group of Kofke.\cite{Shaul12,Kofke19} 
Because the two-body potential used in that work\cite{u2_2010} differs negligibly from the current state of the art for the purpose of calculating $D(T)$, we could consider those calculations to be additional data that supplement ours, making use of their somewhat smaller statistical uncertainties to improve our estimates of the Boltzmann contributions.

The exchange contributions to $D$ for $^4$He are relatively small, due to terms of opposite sign that mostly cancel each other.  Because of this cancellation, our statistical uncertainty for the exchange contribution is similar in magnitude to the contribution itself.  In contrast, for the fermion $^3$He, the exchange contributions all contribute in the positive direction, producing a significant effect on $D(T)$ at temperatures of roughly 4~K and below.

Several factors contribute to the uncertainty of the results.
At low temperatures, the statistical uncertainty of the PIMC calculations is significant, not only in its own right but also in its contribution to the uncertainty $u_V(D)$ due to the uncertainty in the potential, which at the lowest temperatures is estimated by taking differences between PIMC calculations with shifted potentials.
This could be somewhat improved with more computer resources, and also by employing some of the optimized methods for integrating virial coefficients developed by the group of Kofke.\cite{Shaul12,Kofke19}

Uncertainty due to imperfect knowledge of the potential cannot be reduced simply with more computer time; this aspect, which dominates the uncertainty at higher temperatures, requires careful development of potential-energy surfaces.
The pair potential is now known with extraordinary accuracy;\cite{u2_2020} further improvement may be desirable for other reasons but it will not reduce the uncertainty of virial coefficients beyond $B(T)$.
The three-body potential has an estimated expanded ($k = 2$) uncertainty of 2\% at all configurations,\cite{FCI} which could be reduced somewhat with a concerted effort.
Finally, the four-body potential is unknown, and is the largest uncertainty above roughly 10~K.  Even an approximate four-body nonadditive potential, for example with a 20\% uncertainty, would have a large impact on reducing the uncertainty of $D(T)$.  
The most important aspect of such a surface for metrology near room temperature would be its behavior at short-distance configurations dominated by repulsion, since those are the most important configurations at those temperatures.
For use at cryogenic temperatures where the four-body uncertainty is still significant but the dispersion contribution is likely more important, its long-range behavior should approach that derived by Bade.\cite{Bade3}

\begin{acknowledgments}
The authors thank David Kofke and Andrew Schultz of the University at
Buffalo for useful discussions on the calculation of the Boltzmann part of
$D(T)$, Mark McLinden of NIST for insight into the experiments reported in
Refs. \onlinecite{McLinden_2007} and \onlinecite{Moldover_2010}, Christof Gaiser of the PTB for extracting a value of $D$ from the experiments of Ref. \onlinecite{Gaiser_2019}, and Dan
Friend for suggestions that improved clarity.

G.G. acknowledges support from {\em Real-K} project 18SIB02,
which has received funding from the EMPIR programme co-financed by the
Participating States and from the European Union's Horizon 2020 research and
innovation programme.

The calculations for $T \geq 20$~K have been performed on the computing
cluster KORE at Fondazione Bruno Kessler.
We acknowledge the CINECA award IscraC-FAVOHLA under the ISCRA initiative, for the
availability of high performance computing resources and support.
\end{acknowledgments}

\section*{Data Availability}
The data that support the findings of this study are available within this article.

\bibliography{D-He}

\begin{thebibliography}{44}%
\makeatletter
\providecommand \@ifxundefined [1]{%
 \@ifx{#1\undefined}
}%
\providecommand \@ifnum [1]{%
 \ifnum #1\expandafter \@firstoftwo
 \else \expandafter \@secondoftwo
 \fi
}%
\providecommand \@ifx [1]{%
 \ifx #1\expandafter \@firstoftwo
 \else \expandafter \@secondoftwo
 \fi
}%
\providecommand \natexlab [1]{#1}%
\providecommand \enquote  [1]{``#1''}%
\providecommand \bibnamefont  [1]{#1}%
\providecommand \bibfnamefont [1]{#1}%
\providecommand \citenamefont [1]{#1}%
\providecommand \href@noop [0]{\@secondoftwo}%
\providecommand \href [0]{\begingroup \@sanitize@url \@href}%
\providecommand \@href[1]{\@@startlink{#1}\@@href}%
\providecommand \@@href[1]{\endgroup#1\@@endlink}%
\providecommand \@sanitize@url [0]{\catcode `\\12\catcode `\$12\catcode
  `\&12\catcode `\#12\catcode `\^12\catcode `\_12\catcode `\%12\relax}%
\providecommand \@@startlink[1]{}%
\providecommand \@@endlink[0]{}%
\providecommand \url  [0]{\begingroup\@sanitize@url \@url }%
\providecommand \@url [1]{\endgroup\@href {#1}{\urlprefix }}%
\providecommand \urlprefix  [0]{URL }%
\providecommand \Eprint [0]{\href }%
\providecommand \doibase [0]{http://dx.doi.org/}%
\providecommand \selectlanguage [0]{\@gobble}%
\providecommand \bibinfo  [0]{\@secondoftwo}%
\providecommand \bibfield  [0]{\@secondoftwo}%
\providecommand \translation [1]{[#1]}%
\providecommand \BibitemOpen [0]{}%
\providecommand \bibitemStop [0]{}%
\providecommand \bibitemNoStop [0]{.\EOS\space}%
\providecommand \EOS [0]{\spacefactor3000\relax}%
\providecommand \BibitemShut  [1]{\csname bibitem#1\endcsname}%
\let\auto@bib@innerbib\@empty
\bibitem [{\citenamefont {Moldover}\ \emph {et~al.}(2014)\citenamefont
  {Moldover}, \citenamefont {Gavioso}, \citenamefont {Mehl}, \citenamefont
  {Pitre}, \citenamefont {de~Podesta},\ and\ \citenamefont
  {Zhang}}]{Moldover_2014}%
  \BibitemOpen
  \bibfield  {author} {\bibinfo {author} {\bibfnamefont {M.~R.}\ \bibnamefont
  {Moldover}}, \bibinfo {author} {\bibfnamefont {R.~M.}\ \bibnamefont
  {Gavioso}}, \bibinfo {author} {\bibfnamefont {J.~B.}\ \bibnamefont {Mehl}},
  \bibinfo {author} {\bibfnamefont {L.}~\bibnamefont {Pitre}}, \bibinfo
  {author} {\bibfnamefont {M.}~\bibnamefont {de~Podesta}}, \ and\ \bibinfo
  {author} {\bibfnamefont {J.~T.}\ \bibnamefont {Zhang}},\ }\bibfield  {title}
  {\enquote {\bibinfo {title} {Acoustic gas thermometry},}\ }\href {\doibase
  10.1088/0026-1394/51/1/r1} {\bibfield  {journal} {\bibinfo  {journal}
  {Metrologia}\ }\textbf {\bibinfo {volume} {51}},\ \bibinfo {pages} {R1--R19}
  (\bibinfo {year} {2014})}\BibitemShut {NoStop}%
\bibitem [{\citenamefont {Gaiser}, \citenamefont {Zandt},\ and\ \citenamefont
  {Fellmuth}(2015)}]{Gaiser_2015}%
  \BibitemOpen
  \bibfield  {author} {\bibinfo {author} {\bibfnamefont {C.}~\bibnamefont
  {Gaiser}}, \bibinfo {author} {\bibfnamefont {T.}~\bibnamefont {Zandt}}, \
  and\ \bibinfo {author} {\bibfnamefont {B.}~\bibnamefont {Fellmuth}},\
  }\bibfield  {title} {\enquote {\bibinfo {title} {Dielectric-constant gas
  thermometry},}\ }\href {\doibase 10.1088/0026-1394/52/5/s217} {\bibfield
  {journal} {\bibinfo  {journal} {Metrologia}\ }\textbf {\bibinfo {volume}
  {52}},\ \bibinfo {pages} {S217--S226} (\bibinfo {year} {2015})}\BibitemShut
  {NoStop}%
\bibitem [{\citenamefont {Rourke}\ \emph {et~al.}(2019)\citenamefont {Rourke},
  \citenamefont {Gaiser}, \citenamefont {Gao}, \citenamefont {{Madonna Ripa}},
  \citenamefont {Moldover}, \citenamefont {Pitre},\ and\ \citenamefont
  {Underwood}}]{Rourke_2019}%
  \BibitemOpen
  \bibfield  {author} {\bibinfo {author} {\bibfnamefont {P.~M.~C.}\
  \bibnamefont {Rourke}}, \bibinfo {author} {\bibfnamefont {C.}~\bibnamefont
  {Gaiser}}, \bibinfo {author} {\bibfnamefont {B.}~\bibnamefont {Gao}},
  \bibinfo {author} {\bibfnamefont {D.}~\bibnamefont {{Madonna Ripa}}},
  \bibinfo {author} {\bibfnamefont {M.~R.}\ \bibnamefont {Moldover}}, \bibinfo
  {author} {\bibfnamefont {L.}~\bibnamefont {Pitre}}, \ and\ \bibinfo {author}
  {\bibfnamefont {R.~J.}\ \bibnamefont {Underwood}},\ }\bibfield  {title}
  {\enquote {\bibinfo {title} {Refractive-index gas thermometry},}\ }\href
  {\doibase 10.1088/1681-7575/ab0dbe} {\bibfield  {journal} {\bibinfo
  {journal} {Metrologia}\ }\textbf {\bibinfo {volume} {56}},\ \bibinfo {pages}
  {032001} (\bibinfo {year} {2019})}\BibitemShut {NoStop}%
\bibitem [{\citenamefont {Gaiser}, \citenamefont {Fellmuth},\ and\
  \citenamefont {Sabuga}(2020)}]{Gaiser20}%
  \BibitemOpen
  \bibfield  {author} {\bibinfo {author} {\bibfnamefont {C.}~\bibnamefont
  {Gaiser}}, \bibinfo {author} {\bibfnamefont {B.}~\bibnamefont {Fellmuth}}, \
  and\ \bibinfo {author} {\bibfnamefont {W.}~\bibnamefont {Sabuga}},\
  }\bibfield  {title} {\enquote {\bibinfo {title} {Primary gas-pressure
  standard from electrical measurements and thermophysical ab initio
  calculations.}}\ }\href {\doibase 10.1038/s41567-019-0722-2} {\bibfield
  {journal} {\bibinfo  {journal} {Nature Phys.}\ }\textbf {\bibinfo {volume}
  {16}},\ \bibinfo {pages} {177--180} (\bibinfo {year} {2020})}\BibitemShut
  {NoStop}%
\bibitem [{\citenamefont {Cencek}\ \emph {et~al.}(2012)\citenamefont {Cencek},
  \citenamefont {Przybytek}, \citenamefont {Komasa}, \citenamefont {Mehl},
  \citenamefont {Jeziorski},\ and\ \citenamefont {Szalewicz}}]{u2_2010}%
  \BibitemOpen
  \bibfield  {author} {\bibinfo {author} {\bibfnamefont {W.}~\bibnamefont
  {Cencek}}, \bibinfo {author} {\bibfnamefont {M.}~\bibnamefont {Przybytek}},
  \bibinfo {author} {\bibfnamefont {J.}~\bibnamefont {Komasa}}, \bibinfo
  {author} {\bibfnamefont {J.~B.}\ \bibnamefont {Mehl}}, \bibinfo {author}
  {\bibfnamefont {B.}~\bibnamefont {Jeziorski}}, \ and\ \bibinfo {author}
  {\bibfnamefont {K.}~\bibnamefont {Szalewicz}},\ }\bibfield  {title} {\enquote
  {\bibinfo {title} {Effects of adiabatic, relativistic, and quantum
  electrodynamics interactions on the pair potential and thermophysical
  properties of helium},}\ }\href {\doibase 10.1063/1.4712218} {\bibfield
  {journal} {\bibinfo  {journal} {J. Chem. Phys.}\ }\textbf {\bibinfo {volume}
  {136}},\ \bibinfo {pages} {224303} (\bibinfo {year} {2012})}\BibitemShut
  {NoStop}%
\bibitem [{\citenamefont {Przybytek}\ \emph {et~al.}(2017)\citenamefont
  {Przybytek}, \citenamefont {Cencek}, \citenamefont {Jeziorski},\ and\
  \citenamefont {Szalewicz}}]{Przybytek_2017}%
  \BibitemOpen
  \bibfield  {author} {\bibinfo {author} {\bibfnamefont {M.}~\bibnamefont
  {Przybytek}}, \bibinfo {author} {\bibfnamefont {W.}~\bibnamefont {Cencek}},
  \bibinfo {author} {\bibfnamefont {B.}~\bibnamefont {Jeziorski}}, \ and\
  \bibinfo {author} {\bibfnamefont {K.}~\bibnamefont {Szalewicz}},\ }\bibfield
  {title} {\enquote {\bibinfo {title} {Pair potential with submillikelvin
  uncertainties and nonadiabatic treatment of the halo state of the helium
  dimer},}\ }\href {\doibase 10.1103/PhysRevLett.119.123401} {\bibfield
  {journal} {\bibinfo  {journal} {Phys. Rev. Lett.}\ }\textbf {\bibinfo
  {volume} {119}},\ \bibinfo {pages} {123401} (\bibinfo {year}
  {2017})}\BibitemShut {NoStop}%
\bibitem [{\citenamefont {Czachorowski}\ \emph {et~al.}(2020)\citenamefont
  {Czachorowski}, \citenamefont {Przybytek}, \citenamefont {Lesiuk},
  \citenamefont {Puchalski},\ and\ \citenamefont {Jeziorski}}]{u2_2020}%
  \BibitemOpen
  \bibfield  {author} {\bibinfo {author} {\bibfnamefont {P.}~\bibnamefont
  {Czachorowski}}, \bibinfo {author} {\bibfnamefont {M.}~\bibnamefont
  {Przybytek}}, \bibinfo {author} {\bibfnamefont {M.}~\bibnamefont {Lesiuk}},
  \bibinfo {author} {\bibfnamefont {M.}~\bibnamefont {Puchalski}}, \ and\
  \bibinfo {author} {\bibfnamefont {B.}~\bibnamefont {Jeziorski}},\ }\bibfield
  {title} {\enquote {\bibinfo {title} {Second virial coefficients for $^4${He}
  and $^3${He} from an accurate relativistic interaction potential},}\ }\href
  {\doibase 10.1103/PhysRevA.102.042810} {\bibfield  {journal} {\bibinfo
  {journal} {Phys. Rev. A}\ }\textbf {\bibinfo {volume} {102}},\ \bibinfo
  {pages} {042810} (\bibinfo {year} {2020})}\BibitemShut {NoStop}%
\bibitem [{\citenamefont {Hirschfelder}, \citenamefont {Curtiss},\ and\
  \citenamefont {Bird}(1954)}]{Hirschfelder:54}%
  \BibitemOpen
  \bibfield  {author} {\bibinfo {author} {\bibfnamefont {J.~O.}\ \bibnamefont
  {Hirschfelder}}, \bibinfo {author} {\bibfnamefont {C.~F.}\ \bibnamefont
  {Curtiss}}, \ and\ \bibinfo {author} {\bibfnamefont {R.~B.}\ \bibnamefont
  {Bird}},\ }\href@noop {} {\emph {\bibinfo {title} {Molecular Theory of Gases
  and Liquids}}}\ (\bibinfo  {publisher} {John Wiley \& Sons},\ \bibinfo
  {address} {New York},\ \bibinfo {year} {1954})\BibitemShut {NoStop}%
\bibitem [{\citenamefont {Garberoglio}\ and\ \citenamefont
  {Harvey}(2011)}]{Garberoglio11}%
  \BibitemOpen
  \bibfield  {author} {\bibinfo {author} {\bibfnamefont {G.}~\bibnamefont
  {Garberoglio}}\ and\ \bibinfo {author} {\bibfnamefont {A.~H.}\ \bibnamefont
  {Harvey}},\ }\bibfield  {title} {\enquote {\bibinfo {title} {Path-integral
  calculation of the third virial coefficient of quantum gases at low
  temperatures},}\ }\href@noop {} {\bibfield  {journal} {\bibinfo  {journal}
  {J. Chem. Phys.}\ }\textbf {\bibinfo {volume} {134}},\ \bibinfo {pages}
  {134106} (\bibinfo {year} {2011})}\BibitemShut {NoStop}%
\bibitem [{\citenamefont {Garberoglio}\ and\ \citenamefont
  {Harvey}(2020)}]{Garberoglio11err}%
  \BibitemOpen
  \bibfield  {author} {\bibinfo {author} {\bibfnamefont {G.}~\bibnamefont
  {Garberoglio}}\ and\ \bibinfo {author} {\bibfnamefont {A.~H.}\ \bibnamefont
  {Harvey}},\ }\bibfield  {title} {\enquote {\bibinfo {title} {Erratum:
  {P}ath-integral calculation of the third virial coefficient of quantum gases
  at low temperatures},}\ }\href@noop {} {\bibfield  {journal} {\bibinfo
  {journal} {J. Chem. Phys.}\ }\textbf {\bibinfo {volume} {152}},\ \bibinfo
  {pages} {199903} (\bibinfo {year} {2020})}\BibitemShut {NoStop}%
\bibitem [{\citenamefont {Garberoglio}, \citenamefont {Moldover},\ and\
  \citenamefont {Harvey}(2011)}]{Garberoglio2011}%
  \BibitemOpen
  \bibfield  {author} {\bibinfo {author} {\bibfnamefont {G.}~\bibnamefont
  {Garberoglio}}, \bibinfo {author} {\bibfnamefont {M.~R.}\ \bibnamefont
  {Moldover}}, \ and\ \bibinfo {author} {\bibfnamefont {A.~H.}\ \bibnamefont
  {Harvey}},\ }\bibfield  {title} {\enquote {\bibinfo {title} {Improved
  first-principles calculation of the third virial coefficient of helium},}\
  }\href@noop {} {\bibfield  {journal} {\bibinfo  {journal} {J. Res. Natl.
  Inst. Stand. Technol.}\ }\textbf {\bibinfo {volume} {116}},\ \bibinfo {pages}
  {729--742} (\bibinfo {year} {2011})}\BibitemShut {NoStop}%
\bibitem [{\citenamefont {Garberoglio}, \citenamefont {Moldover},\ and\
  \citenamefont {Harvey}(2020)}]{improvederr}%
  \BibitemOpen
  \bibfield  {author} {\bibinfo {author} {\bibfnamefont {G.}~\bibnamefont
  {Garberoglio}}, \bibinfo {author} {\bibfnamefont {M.~R.}\ \bibnamefont
  {Moldover}}, \ and\ \bibinfo {author} {\bibfnamefont {A.~H.}\ \bibnamefont
  {Harvey}},\ }\bibfield  {title} {\enquote {\bibinfo {title} {Erratum:
  Improved first-principles calculation of the third virial coefficient of
  helium},}\ }\href@noop {} {\bibfield  {journal} {\bibinfo  {journal} {J. Res.
  Natl. Inst. Stand. Technol.}\ }\textbf {\bibinfo {volume} {125}},\ \bibinfo
  {pages} {125019} (\bibinfo {year} {2020})}\BibitemShut {NoStop}%
\bibitem [{\citenamefont {Cencek}, \citenamefont {Patkowski},\ and\
  \citenamefont {Szalewicz}(2009)}]{FCI}%
  \BibitemOpen
  \bibfield  {author} {\bibinfo {author} {\bibfnamefont {W.}~\bibnamefont
  {Cencek}}, \bibinfo {author} {\bibfnamefont {K.}~\bibnamefont {Patkowski}}, \
  and\ \bibinfo {author} {\bibfnamefont {K.}~\bibnamefont {Szalewicz}},\
  }\bibfield  {title} {\enquote {\bibinfo {title}
  {Full-configuration-interaction calculation of three-body nonadditive
  contribution to helium interaction potential},}\ }\href@noop {} {\bibfield
  {journal} {\bibinfo  {journal} {J. Chem. Phys.}\ }\textbf {\bibinfo {volume}
  {131}},\ \bibinfo {pages} {064105} (\bibinfo {year} {2009})}\BibitemShut
  {NoStop}%
\bibitem [{\citenamefont {Shaul}, \citenamefont {Schultz},\ and\ \citenamefont
  {Kofke}(2012)}]{Shaul12}%
  \BibitemOpen
  \bibfield  {author} {\bibinfo {author} {\bibfnamefont {K.~R.~S.}\
  \bibnamefont {Shaul}}, \bibinfo {author} {\bibfnamefont {A.~J.}\ \bibnamefont
  {Schultz}}, \ and\ \bibinfo {author} {\bibfnamefont {D.~A.}\ \bibnamefont
  {Kofke}},\ }\bibfield  {title} {\enquote {\bibinfo {title} {Path-integral
  {M}ayer-sampling calculations of the quantum {B}oltzmann contribution to
  virial coefficients of helium-4},}\ }\href@noop {} {\bibfield  {journal}
  {\bibinfo  {journal} {J. Chem. Phys.}\ }\textbf {\bibinfo {volume} {137}},\
  \bibinfo {pages} {184101} (\bibinfo {year} {2012})}\BibitemShut {NoStop}%
\bibitem [{\citenamefont {Schultz}\ and\ \citenamefont
  {Kofke}(2019)}]{Kofke19}%
  \BibitemOpen
  \bibfield  {author} {\bibinfo {author} {\bibfnamefont {A.~J.}\ \bibnamefont
  {Schultz}}\ and\ \bibinfo {author} {\bibfnamefont {D.~A.}\ \bibnamefont
  {Kofke}},\ }\bibfield  {title} {\enquote {\bibinfo {title} {Virial
  coefficients of helium-4 from \textit{ab initio}-based molecular models},}\
  }\href@noop {} {\bibfield  {journal} {\bibinfo  {journal} {J. Chem. Eng.
  Data}\ }\textbf {\bibinfo {volume} {64}},\ \bibinfo {pages} {3742--3754}
  (\bibinfo {year} {2019})}\BibitemShut {NoStop}%
\bibitem [{\citenamefont {Gao}\ \emph {et~al.}(2020)\citenamefont {Gao},
  \citenamefont {Zhang}, \citenamefont {Han}, \citenamefont {Pan},
  \citenamefont {Chen}, \citenamefont {Song}, \citenamefont {Liu},
  \citenamefont {Hu}, \citenamefont {Kong}, \citenamefont {Sparasci},
  \citenamefont {Plimmer}, \citenamefont {Luo},\ and\ \citenamefont
  {Pitre}}]{Gao_2020}%
  \BibitemOpen
  \bibfield  {author} {\bibinfo {author} {\bibfnamefont {B.}~\bibnamefont
  {Gao}}, \bibinfo {author} {\bibfnamefont {H.}~\bibnamefont {Zhang}}, \bibinfo
  {author} {\bibfnamefont {D.}~\bibnamefont {Han}}, \bibinfo {author}
  {\bibfnamefont {C.}~\bibnamefont {Pan}}, \bibinfo {author} {\bibfnamefont
  {H.}~\bibnamefont {Chen}}, \bibinfo {author} {\bibfnamefont {Y.}~\bibnamefont
  {Song}}, \bibinfo {author} {\bibfnamefont {W.}~\bibnamefont {Liu}}, \bibinfo
  {author} {\bibfnamefont {J.}~\bibnamefont {Hu}}, \bibinfo {author}
  {\bibfnamefont {X.}~\bibnamefont {Kong}}, \bibinfo {author} {\bibfnamefont
  {F.}~\bibnamefont {Sparasci}}, \bibinfo {author} {\bibfnamefont
  {M.}~\bibnamefont {Plimmer}}, \bibinfo {author} {\bibfnamefont
  {E.}~\bibnamefont {Luo}}, \ and\ \bibinfo {author} {\bibfnamefont
  {L.}~\bibnamefont {Pitre}},\ }\bibfield  {title} {\enquote {\bibinfo {title}
  {Measurement of thermodynamic temperature between 5 {K} and 24.5 {K} with
  single-pressure refractive-index gas thermometry},}\ }\href {\doibase
  10.1088/1681-7575/ab84ca} {\bibfield  {journal} {\bibinfo  {journal}
  {Metrologia}\ }\textbf {\bibinfo {volume} {57}},\ \bibinfo {pages} {065006}
  (\bibinfo {year} {2020})}\BibitemShut {NoStop}%
\bibitem [{\citenamefont {Garberoglio}(2012)}]{Garberoglio2012}%
  \BibitemOpen
  \bibfield  {author} {\bibinfo {author} {\bibfnamefont {G.}~\bibnamefont
  {Garberoglio}},\ }\bibfield  {title} {\enquote {\bibinfo {title} {Quantum
  effects on virial coefficients: a numerical approach using centroids},}\
  }\href@noop {} {\bibfield  {journal} {\bibinfo  {journal} {Chem. Phys.
  Lett.}\ }\textbf {\bibinfo {volume} {525-526}},\ \bibinfo {pages} {19--23}
  (\bibinfo {year} {2012})}\BibitemShut {NoStop}%
\bibitem [{\citenamefont {Hill}(1987{\natexlab{a}})}]{Hill-intro}%
  \BibitemOpen
  \bibfield  {author} {\bibinfo {author} {\bibfnamefont {T.~L.}\ \bibnamefont
  {Hill}},\ }\href@noop {} {\emph {\bibinfo {title} {An Introduction to
  Statistical Thermodynamics}}}\ (\bibinfo  {publisher} {Dover},\ \bibinfo
  {address} {New York},\ \bibinfo {year} {1987})\BibitemShut {NoStop}%
\bibitem [{\citenamefont {Hill}(1987{\natexlab{b}})}]{Hill}%
  \BibitemOpen
  \bibfield  {author} {\bibinfo {author} {\bibfnamefont {T.~L.}\ \bibnamefont
  {Hill}},\ }\href@noop {} {\emph {\bibinfo {title} {Statistical Mechanics}}}\
  (\bibinfo  {publisher} {Dover},\ \bibinfo {address} {New York},\ \bibinfo
  {year} {1987})\BibitemShut {NoStop}%
\bibitem [{\citenamefont {Boyd}, \citenamefont {Larsen},\ and\ \citenamefont
  {Kilpatrick}(1969)}]{Boyd69}%
  \BibitemOpen
  \bibfield  {author} {\bibinfo {author} {\bibfnamefont {M.~E.}\ \bibnamefont
  {Boyd}}, \bibinfo {author} {\bibfnamefont {S.~Y.}\ \bibnamefont {Larsen}}, \
  and\ \bibinfo {author} {\bibfnamefont {J.~E.}\ \bibnamefont {Kilpatrick}},\
  }\bibfield  {title} {\enquote {\bibinfo {title} {Quantum mechanical second
  virial coefficient of a {L}ennard-{J}ones gas. {H}elium},}\ }\href@noop {}
  {\bibfield  {journal} {\bibinfo  {journal} {J. Chem. Phys.}\ }\textbf
  {\bibinfo {volume} {50}},\ \bibinfo {pages} {4034--4055} (\bibinfo {year}
  {1969})}\BibitemShut {NoStop}%
\bibitem [{\citenamefont {Lepage}(1980)}]{vegas2}%
  \BibitemOpen
  \bibfield  {author} {\bibinfo {author} {\bibfnamefont {G.}~\bibnamefont
  {Lepage}},\ }\href@noop {} {\enquote {\bibinfo {title} {{VEGAS}: An adaptive
  multi-dimensional integration program},}\ }\bibinfo {type} {Tech. Rep.}\
  (\bibinfo  {institution} {Cornell preprint CLNS 80-447},\ \bibinfo {year}
  {1980})\BibitemShut {NoStop}%
\bibitem [{\citenamefont {Kreckel}(1997)}]{Kreckel97}%
  \BibitemOpen
  \bibfield  {author} {\bibinfo {author} {\bibfnamefont {R.}~\bibnamefont
  {Kreckel}},\ }\bibfield  {title} {\enquote {\bibinfo {title} {Parallelization
  of adaptive {MC} integrators},}\ }\href@noop {} {\bibfield  {journal}
  {\bibinfo  {journal} {Comput. Phys. Commun.}\ }\textbf {\bibinfo {volume}
  {106}},\ \bibinfo {pages} {258--266} (\bibinfo {year} {1997})}\BibitemShut
  {NoStop}%
\bibitem [{\citenamefont {Levy}(1954)}]{Levy54}%
  \BibitemOpen
  \bibfield  {author} {\bibinfo {author} {\bibfnamefont {P.}~\bibnamefont
  {Levy}},\ }\href@noop {} {\emph {\bibinfo {title} {Memorial des Sciences
  Mathematiques}}}\ (\bibinfo  {publisher} {Gauthier Villars},\ \bibinfo
  {address} {Paris},\ \bibinfo {year} {1954})\ \bibinfo {note} {fascicule
  126}\BibitemShut {NoStop}%
\bibitem [{\citenamefont {Jordan}\ and\ \citenamefont
  {Fosdick}(1968)}]{Jordan-Fosdick68}%
  \BibitemOpen
  \bibfield  {author} {\bibinfo {author} {\bibfnamefont {H.~F.}\ \bibnamefont
  {Jordan}}\ and\ \bibinfo {author} {\bibfnamefont {L.~D.}\ \bibnamefont
  {Fosdick}},\ }\bibfield  {title} {\enquote {\bibinfo {title} {Three-particle
  effects in the pair distribution function for {H}e${}^4$ gas},}\ }\href@noop
  {} {\bibfield  {journal} {\bibinfo  {journal} {Phys. Rev.}\ }\textbf
  {\bibinfo {volume} {171}},\ \bibinfo {pages} {128--149} (\bibinfo {year}
  {1968})}\BibitemShut {NoStop}%
\bibitem [{\citenamefont {Garberoglio}\ and\ \citenamefont
  {Harvey}(2009)}]{Garberoglio2009b}%
  \BibitemOpen
  \bibfield  {author} {\bibinfo {author} {\bibfnamefont {G.}~\bibnamefont
  {Garberoglio}}\ and\ \bibinfo {author} {\bibfnamefont {A.~H.}\ \bibnamefont
  {Harvey}},\ }\bibfield  {title} {\enquote {\bibinfo {title} {First-principles
  calculation of the third virial coefficient of helium},}\ }\href@noop {}
  {\bibfield  {journal} {\bibinfo  {journal} {J. Res. Natl. Inst. Stand.
  Technol.}\ }\textbf {\bibinfo {volume} {114}},\ \bibinfo {pages} {249--262}
  (\bibinfo {year} {2009})}\BibitemShut {NoStop}%
\bibitem [{\citenamefont {Cencek}\ \emph {et~al.}(2013)\citenamefont {Cencek},
  \citenamefont {Garberoglio}, \citenamefont {Harvey}, \citenamefont
  {McLinden},\ and\ \citenamefont {Szalewicz}}]{Cencek2013}%
  \BibitemOpen
  \bibfield  {author} {\bibinfo {author} {\bibfnamefont {W.}~\bibnamefont
  {Cencek}}, \bibinfo {author} {\bibfnamefont {G.}~\bibnamefont {Garberoglio}},
  \bibinfo {author} {\bibfnamefont {A.~H.}\ \bibnamefont {Harvey}}, \bibinfo
  {author} {\bibfnamefont {M.~O.}\ \bibnamefont {McLinden}}, \ and\ \bibinfo
  {author} {\bibfnamefont {K.}~\bibnamefont {Szalewicz}},\ }\bibfield  {title}
  {\enquote {\bibinfo {title} {Three-body nonadditive potential for argon with
  estimated uncertainties and third virial coefficient},}\ }\href@noop {}
  {\bibfield  {journal} {\bibinfo  {journal} {J. Phys. Chem. A}\ }\textbf
  {\bibinfo {volume} {117}},\ \bibinfo {pages} {7542--7552} (\bibinfo {year}
  {2013})}\BibitemShut {NoStop}%
\bibitem [{\citenamefont {Rodr{\'\i}guez-Cantano}\ \emph
  {et~al.}(2016)\citenamefont {Rodr{\'\i}guez-Cantano}, \citenamefont
  {P{\'e}rez~de Tudela}, \citenamefont {Bartolomei}, \citenamefont
  {Hern{\'a}ndez}, \citenamefont {Campos-Mart{\'\i}nez}, \citenamefont
  {Gonz{\'a}lez-Lezana}, \citenamefont {Villarreal}, \citenamefont
  {Hern{\'a}ndez-Rojas},\ and\ \citenamefont {Bretón}}]{FH4}%
  \BibitemOpen
  \bibfield  {author} {\bibinfo {author} {\bibfnamefont {R.}~\bibnamefont
  {Rodr{\'\i}guez-Cantano}}, \bibinfo {author} {\bibfnamefont {R.}~\bibnamefont
  {P{\'e}rez~de Tudela}}, \bibinfo {author} {\bibfnamefont {M.}~\bibnamefont
  {Bartolomei}}, \bibinfo {author} {\bibfnamefont {M.~I.}\ \bibnamefont
  {Hern{\'a}ndez}}, \bibinfo {author} {\bibfnamefont {J.}~\bibnamefont
  {Campos-Mart{\'\i}nez}}, \bibinfo {author} {\bibfnamefont {T.}~\bibnamefont
  {Gonz{\'a}lez-Lezana}}, \bibinfo {author} {\bibfnamefont {P.}~\bibnamefont
  {Villarreal}}, \bibinfo {author} {\bibfnamefont {J.}~\bibnamefont
  {Hern{\'a}ndez-Rojas}}, \ and\ \bibinfo {author} {\bibfnamefont
  {J.}~\bibnamefont {Bretón}},\ }\bibfield  {title} {\enquote {\bibinfo
  {title} {Examination of the {F}eynman--{H}ibbs approach in the study of
  {N}e{N}-coronene clusters at low temperatures},}\ }\href@noop {} {\bibfield
  {journal} {\bibinfo  {journal} {J. Phys. Chem. A}\ }\textbf {\bibinfo
  {volume} {120}},\ \bibinfo {pages} {5370--5379} (\bibinfo {year}
  {2016})}\BibitemShut {NoStop}%
\bibitem [{\citenamefont {Bade}(1957)}]{Bade1}%
  \BibitemOpen
  \bibfield  {author} {\bibinfo {author} {\bibfnamefont {W.~L.}\ \bibnamefont
  {Bade}},\ }\bibfield  {title} {\enquote {\bibinfo {title} {Drude-model
  calculation of dispersion forces. {I}. {G}eneral theory},}\ }\href@noop {}
  {\bibfield  {journal} {\bibinfo  {journal} {J. Chem. Phys.}\ }\textbf
  {\bibinfo {volume} {27}},\ \bibinfo {pages} {1280--1284} (\bibinfo {year}
  {1957})}\BibitemShut {NoStop}%
\bibitem [{\citenamefont {Bade}(1958)}]{Bade3}%
  \BibitemOpen
  \bibfield  {author} {\bibinfo {author} {\bibfnamefont {W.~L.}\ \bibnamefont
  {Bade}},\ }\bibfield  {title} {\enquote {\bibinfo {title} {Drude-model
  calculation of dispersion forces. {III}. {T}he fourth-order contribution},}\
  }\href@noop {} {\bibfield  {journal} {\bibinfo  {journal} {J. Chem. Phys.}\
  }\textbf {\bibinfo {volume} {28}},\ \bibinfo {pages} {282--284} (\bibinfo
  {year} {1958})}\BibitemShut {NoStop}%
\bibitem [{\citenamefont {Wiebke}, \citenamefont {Pahl},\ and\ \citenamefont
  {Schwerdtfeger}(2012{\natexlab{a}})}]{WiebkeNe}%
  \BibitemOpen
  \bibfield  {author} {\bibinfo {author} {\bibfnamefont {J.}~\bibnamefont
  {Wiebke}}, \bibinfo {author} {\bibfnamefont {E.}~\bibnamefont {Pahl}}, \ and\
  \bibinfo {author} {\bibfnamefont {P.}~\bibnamefont {Schwerdtfeger}},\
  }\bibfield  {title} {\enquote {\bibinfo {title} {Up to fourth virial
  coefficients from simple and efficient internal-coordinate sampling:
  Application to neon},}\ }\href@noop {} {\bibfield  {journal} {\bibinfo
  {journal} {J. Chem. Phys.}\ }\textbf {\bibinfo {volume} {137}},\ \bibinfo
  {pages} {014508} (\bibinfo {year} {2012}{\natexlab{a}})}\BibitemShut
  {NoStop}%
\bibitem [{\citenamefont {Wiebke}, \citenamefont {Pahl},\ and\ \citenamefont
  {Schwerdtfeger}(2012{\natexlab{b}})}]{WiebkeAr}%
  \BibitemOpen
  \bibfield  {author} {\bibinfo {author} {\bibfnamefont {J.}~\bibnamefont
  {Wiebke}}, \bibinfo {author} {\bibfnamefont {E.}~\bibnamefont {Pahl}}, \ and\
  \bibinfo {author} {\bibfnamefont {P.}~\bibnamefont {Schwerdtfeger}},\
  }\bibfield  {title} {\enquote {\bibinfo {title} {Sensitivity of the thermal
  and acoustic virial coefficients of argon to the argon interaction
  potential},}\ }\href@noop {} {\bibfield  {journal} {\bibinfo  {journal} {J.
  Chem. Phys.}\ }\textbf {\bibinfo {volume} {137}},\ \bibinfo {pages} {064702}
  (\bibinfo {year} {2012}{\natexlab{b}})}\BibitemShut {NoStop}%
\bibitem [{\citenamefont {Gaiser}\ and\ \citenamefont
  {Fellmuth}(2018)}]{Gaiser18}%
  \BibitemOpen
  \bibfield  {author} {\bibinfo {author} {\bibfnamefont {C.}~\bibnamefont
  {Gaiser}}\ and\ \bibinfo {author} {\bibfnamefont {B.}~\bibnamefont
  {Fellmuth}},\ }\bibfield  {title} {\enquote {\bibinfo {title} {Polarizability
  of helium, neon, and argon: New perspectives for gas metrology},}\ }\href
  {\doibase 10.1103/PhysRevLett.120.123203} {\bibfield  {journal} {\bibinfo
  {journal} {Phys. Rev. Lett.}\ }\textbf {\bibinfo {volume} {120}},\ \bibinfo
  {pages} {123203} (\bibinfo {year} {2018})}\BibitemShut {NoStop}%
\bibitem [{\citenamefont {Puchalski}\ \emph {et~al.}(2020)\citenamefont
  {Puchalski}, \citenamefont {Szalewicz}, \citenamefont {Lesiuk},\ and\
  \citenamefont {Jeziorski}}]{Puchalski20}%
  \BibitemOpen
  \bibfield  {author} {\bibinfo {author} {\bibfnamefont {M.}~\bibnamefont
  {Puchalski}}, \bibinfo {author} {\bibfnamefont {K.}~\bibnamefont
  {Szalewicz}}, \bibinfo {author} {\bibfnamefont {M.}~\bibnamefont {Lesiuk}}, \
  and\ \bibinfo {author} {\bibfnamefont {B.}~\bibnamefont {Jeziorski}},\
  }\bibfield  {title} {\enquote {\bibinfo {title} {{QED} calculation of the
  dipole polarizability of helium atom},}\ }\href {\doibase
  10.1103/PhysRevA.101.022505} {\bibfield  {journal} {\bibinfo  {journal}
  {Phys. Rev. A}\ }\textbf {\bibinfo {volume} {101}},\ \bibinfo {pages}
  {022505} (\bibinfo {year} {2020})}\BibitemShut {NoStop}%
\bibitem [{\citenamefont {Thakkar}(1981)}]{Thakkar_1981}%
  \BibitemOpen
  \bibfield  {author} {\bibinfo {author} {\bibfnamefont {A.~J.}\ \bibnamefont
  {Thakkar}},\ }\bibfield  {title} {\enquote {\bibinfo {title} {The generator
  coordinate method applied to variational perturbation theory. {M}ultipole
  polarizabilities, spectral sums, and dispersion coefficients for helium},}\
  }\href {\doibase 10.1063/1.442617} {\bibfield  {journal} {\bibinfo  {journal}
  {J. Chem. Phys.}\ }\textbf {\bibinfo {volume} {75}},\ \bibinfo {pages}
  {4496--4501} (\bibinfo {year} {1981})}\BibitemShut {NoStop}%
\bibitem [{\citenamefont {Blancett}, \citenamefont {Hall},\ and\ \citenamefont
  {Canfield}(1970)}]{Blancett_1970}%
  \BibitemOpen
  \bibfield  {author} {\bibinfo {author} {\bibfnamefont {A.~L.}\ \bibnamefont
  {Blancett}}, \bibinfo {author} {\bibfnamefont {K.~R.}\ \bibnamefont {Hall}},
  \ and\ \bibinfo {author} {\bibfnamefont {F.~B.}\ \bibnamefont {Canfield}},\
  }\bibfield  {title} {\enquote {\bibinfo {title} {Isotherms for the {He--Ar}
  system at 50${}^\circ${C}, 0${}^\circ${C} and -50${}^\circ${C} up to 700
  atm},}\ }\href {\doibase 10.1016/0031-8914(70)90101-1} {\bibfield  {journal}
  {\bibinfo  {journal} {Physica}\ }\textbf {\bibinfo {volume} {47}},\ \bibinfo
  {pages} {75--91} (\bibinfo {year} {1970})}\BibitemShut {NoStop}%
\bibitem [{\citenamefont {Hall}\ and\ \citenamefont
  {Canfield}(1970{\natexlab{a}})}]{Hall_1970a}%
  \BibitemOpen
  \bibfield  {author} {\bibinfo {author} {\bibfnamefont {K.~R.}\ \bibnamefont
  {Hall}}\ and\ \bibinfo {author} {\bibfnamefont {F.~B.}\ \bibnamefont
  {Canfield}},\ }\bibfield  {title} {\enquote {\bibinfo {title} {A
  least-squares method for reduction of {B}urnett data to compressibility
  factors and virial coefficients},}\ }\href {\doibase
  10.1016/0031-8914(70)90103-5} {\bibfield  {journal} {\bibinfo  {journal}
  {Physica}\ }\textbf {\bibinfo {volume} {47}},\ \bibinfo {pages} {99--108}
  (\bibinfo {year} {1970}{\natexlab{a}})}\BibitemShut {NoStop}%
\bibitem [{\citenamefont {Hall}\ and\ \citenamefont
  {Canfield}(1970{\natexlab{b}})}]{Hall_1970b}%
  \BibitemOpen
  \bibfield  {author} {\bibinfo {author} {\bibfnamefont {K.~R.}\ \bibnamefont
  {Hall}}\ and\ \bibinfo {author} {\bibfnamefont {F.~B.}\ \bibnamefont
  {Canfield}},\ }\bibfield  {title} {\enquote {\bibinfo {title} {Isotherms for
  the {He--N$_2$} system at -190${}^\circ${C}, -170${}^\circ${C} and
  -160${}^\circ${C} up to 700 atm},}\ }\href {\doibase
  10.1016/0031-8914(70)90281-8} {\bibfield  {journal} {\bibinfo  {journal}
  {Physica}\ }\textbf {\bibinfo {volume} {47}},\ \bibinfo {pages} {219--226}
  (\bibinfo {year} {1970}{\natexlab{b}})}\BibitemShut {NoStop}%
\bibitem [{\citenamefont {Provine}\ and\ \citenamefont
  {Canfield}(1971)}]{Provine_1971}%
  \BibitemOpen
  \bibfield  {author} {\bibinfo {author} {\bibfnamefont {J.~A.}\ \bibnamefont
  {Provine}}\ and\ \bibinfo {author} {\bibfnamefont {F.~B.}\ \bibnamefont
  {Canfield}},\ }\bibfield  {title} {\enquote {\bibinfo {title} {Isotherms for
  the {He--Ar} system at -130, -115, and -90${}^\circ${C} up to 700 atm},}\
  }\href {\doibase 10.1016/0031-8914(71)90038-3} {\bibfield  {journal}
  {\bibinfo  {journal} {Physica}\ }\textbf {\bibinfo {volume} {52}},\ \bibinfo
  {pages} {79--91} (\bibinfo {year} {1971})}\BibitemShut {NoStop}%
\bibitem [{\citenamefont {McLinden}\ and\ \citenamefont
  {Lösch-Will}(2007)}]{McLinden_2007}%
  \BibitemOpen
  \bibfield  {author} {\bibinfo {author} {\bibfnamefont {M.~O.}\ \bibnamefont
  {McLinden}}\ and\ \bibinfo {author} {\bibfnamefont {C.}~\bibnamefont
  {Lösch-Will}},\ }\bibfield  {title} {\enquote {\bibinfo {title} {Apparatus
  for wide-ranging, high-accuracy fluid ($p,\rho,{T}$) measurements based on a
  compact two-sinker densimeter},}\ }\href {\doibase 10.1016/j.jct.2006.09.012}
  {\bibfield  {journal} {\bibinfo  {journal} {J. Chem. Thermodyn.}\ }\textbf
  {\bibinfo {volume} {39}},\ \bibinfo {pages} {507--530} (\bibinfo {year}
  {2007})}\BibitemShut {NoStop}%
\bibitem [{\citenamefont {Moldover}\ and\ \citenamefont
  {McLinden}(2010)}]{Moldover_2010}%
  \BibitemOpen
  \bibfield  {author} {\bibinfo {author} {\bibfnamefont {M.~R.}\ \bibnamefont
  {Moldover}}\ and\ \bibinfo {author} {\bibfnamefont {M.~O.}\ \bibnamefont
  {McLinden}},\ }\bibfield  {title} {\enquote {\bibinfo {title} {Using
  \textit{ab initio} ``data'' to accurately determine the fourth density virial
  coefficient of helium},}\ }\href {\doibase 10.1016/j.jct.2010.02.015}
  {\bibfield  {journal} {\bibinfo  {journal} {J. Chem. Thermodyn.}\ }\textbf
  {\bibinfo {volume} {42}},\ \bibinfo {pages} {1193--1203} (\bibinfo {year}
  {2010})}\BibitemShut {NoStop}%
\bibitem [{\citenamefont {Gaiser}\ and\ \citenamefont
  {Fellmuth}(2019)}]{Gaiser_2019}%
  \BibitemOpen
  \bibfield  {author} {\bibinfo {author} {\bibfnamefont {C.}~\bibnamefont
  {Gaiser}}\ and\ \bibinfo {author} {\bibfnamefont {B.}~\bibnamefont
  {Fellmuth}},\ }\bibfield  {title} {\enquote {\bibinfo {title}
  {Highly-accurate density-virial-coefficient values for helium, neon, and
  argon at 0.01~${}^\circ${C} determined by dielectric-constant gas
  thermometry},}\ }\href {\doibase 10.1063/1.5090224} {\bibfield  {journal}
  {\bibinfo  {journal} {J. Chem. Phys.}\ }\textbf {\bibinfo {volume} {150}},\
  \bibinfo {pages} {134303} (\bibinfo {year} {2019})}\BibitemShut {NoStop}%
\bibitem [{\citenamefont {Gaiser}()}]{Gaiser_2021}%
  \BibitemOpen
  \bibfield  {author} {\bibinfo {author} {\bibfnamefont {C.}~\bibnamefont
  {Gaiser}},\ }\href@noop {} {}\bibinfo {howpublished} {Physikalisch-Technische
  Bundesanstalt},\ \bibinfo {note} {personal communication (2021)}\BibitemShut
  {NoStop}%
\bibitem [{\citenamefont {Shaul}\ \emph {et~al.}(2012)\citenamefont {Shaul},
  \citenamefont {Schultz}, \citenamefont {Kofke},\ and\ \citenamefont
  {Moldover}}]{Shaul_2012}%
  \BibitemOpen
  \bibfield  {author} {\bibinfo {author} {\bibfnamefont {K.~R.~S.}\
  \bibnamefont {Shaul}}, \bibinfo {author} {\bibfnamefont {A.~J.}\ \bibnamefont
  {Schultz}}, \bibinfo {author} {\bibfnamefont {D.~A.}\ \bibnamefont {Kofke}},
  \ and\ \bibinfo {author} {\bibfnamefont {M.~R.}\ \bibnamefont {Moldover}},\
  }\bibfield  {title} {\enquote {\bibinfo {title} {Semiclassical fifth virial
  coefficients for improved \textit{ab initio} helium-4 standards},}\ }\href
  {\doibase https://doi.org/10.1016/j.cplett.2012.02.013} {\bibfield  {journal}
  {\bibinfo  {journal} {Chem. Phys. Lett.}\ }\textbf {\bibinfo {volume}
  {531}},\ \bibinfo {pages} {11--17} (\bibinfo {year} {2012})}\BibitemShut
  {NoStop}%
\bibitem [{\citenamefont {{McLinden}}()}]{McLinden_2020}%
  \BibitemOpen
  \bibfield  {author} {\bibinfo {author} {\bibfnamefont {M.~O.}\ \bibnamefont
  {{McLinden}}},\ }\href@noop {} {}\bibinfo {howpublished} {National Institute
  of Standards and Technology},\ \bibinfo {note} {personal communication
  (2020)}\BibitemShut {NoStop}%
\end{thebibliography}%

\end{document}